\documentclass{aa}  
\usepackage{color}
\usepackage{graphicx}
\usepackage{lscape}
\usepackage{natbib}
\usepackage{natbib,twoopt}
\usepackage[varg]{txfonts}
\usepackage{url}
\usepackage{xspace}
\usepackage{amssymb}
\usepackage{stmaryrd}
\usepackage{siunitx}
\usepackage{textcomp}

\bibpunct{(}{)}{;}{a}{}{,}

\def\etal{{et\,al.}\ }

\newcommand{\jl}{\object{JL\,9}\xspace}
\newcommand{\rl}{\object{RL\,104}\xspace}
\newcommand{\sdss}{\object{SDSS\,J141812.51$-$024426.8}\xspace}
\newcommand{\sdsss}{\object{J1418}\xspace}
\newcommand{\gaiaf}{\object{Gaia DR2 5340389724892633984}\xspace}
\newcommand{\gaias}{\object{Gaia DR2 6735687985029707136}\xspace}
\newcommand{\gaiaff}{\object{Gaia DR2 53}\xspace}
\newcommand{\gaiass}{\object{Gaia DR2 67}\xspace}
\newcounter{Rco}

\newcommand{\logg}{\mbox{$\log g$}\xspace}

\newcommand{\Teff}{\mbox{$T_\mathrm{eff}$}\xspace}

\newcommand{\ebv}{$E_\mathrm{B-V}$\xspace}

\newcommand{\vrad}{$v_\mathrm{rad}$\xspace}

\newcommand{\Lsol}{$L_\odot$}
\newcommand{\Msol}{$M_\odot$}
\newcommand{\Rsol}{$R_\odot$}

\begin{document}

\title{Non-local thermodynamic equilibrium spectral analysis of five hot, hydrogen-deficient pre-white dwarfs
}

\author{Klaus Werner\inst{1} \and Nicole Reindl\inst{2} \and Matti Dorsch\inst{3} \and
  Stephan Geier\inst{2} \and Ulisse Munari\inst{4} \and Roberto Raddi\inst{5}}

\institute{Institut f\"ur Astronomie und Astrophysik, Kepler Center for
  Astro and Particle Physics, Eberhard Karls Universit\"at, Sand~1, 72076
  T\"ubingen, Germany\\ \email{werner@astro.uni-tuebingen.de} 
\and
  Institut f\"ur Physik und Astronomie, Universit\"at Potsdam, Karl-Liebknecht-Stra\ss e 24/25, 14476, Potsdam, Germany
\and
Dr.\ Karl Remeis-Observatory \& ECAP, FAU Erlangen-N\"{u}rnberg, Sternwartstr.\ 7, 96049 Bamberg, Germany 
\and
INAF Astronomical Observatory of Padova, 36012 Asiago (VI), Italy
\and
Departament
de F\'isica, Universitat Polit\`ecnica de Catalunya, c/Esteve Terrades 5, E-08860 Castelldefels, Spain
}

\date{xx xx 2021 / xx xx 2021}

\authorrunning{K. Werner \etal}
\titlerunning{Non-LTE spectral analysis of five hot hydrogen-deficient pre-white dwarfs}

\abstract{Hot, compact, hydrogen-deficient pre-white dwarfs (pre-WDs)
  with effective temperatures of \Teff $>$ 70\,000\,K and a surface gravity of
  5.0 $<$ \logg $<$ 7.0 are rather rare objects despite recent and
  ongoing surveys. It is believed that they are the outcome of either
  single star evolution (late helium-shell flash or late helium-core
  flash) or binary star evolution (double WD merger). Their study is
  interesting because the surface elemental abundances reflect the
  physics of thermonuclear flashes and merger
  events. Spectroscopically they are divided in three different
  classes, namely PG1159, O(He), or He-sdO. We present a
  spectroscopic analysis of five such stars that turned out to have
  atmospheric parameters in the range \Teff = 70\,000--80\,000\,K and
  \logg = 5.2--6.3. The three investigated He-sdOs have a relatively
  high hydrogen mass fraction (10\%) that is unexplained by both
  single (He core flash) and binary evolution (He-WD merger)
  scenarios. The O(He) star \jl is probably a binary helium-WD merger,
  but its hydrogen content (6\%) is also at odds with merger
  models. We found that \rl is the `coolest' (\Teff =
  80\,000\,K) member of the PG1159 class in a pre-WD
  stage. Its optical spectrum is remarkable because it exhibits
  \ion{C}{iv} lines involving Rydberg states with principal quantum
  numbers up to $n=22$. Its rather low mass ($0.48^{+0.03}_{-0.02}
  M_\odot$) is difficult to reconcile with the common evolutionary
  scenario for PG1159 stars due to it being the outcome of a (very) late
  He-shell flash. The same mass-problem faces a merger model of a
  close He-sdO plus CO WD binary that predicts PG1159-like
  abundances. Perhaps \rl originates from a very late He-shell flash
  in a CO/He WD formed by a merger of two low-mass He-WDs.}


\maketitle
%

\begin{figure*}[t]
 \centering  \includegraphics[width=0.9\textwidth]{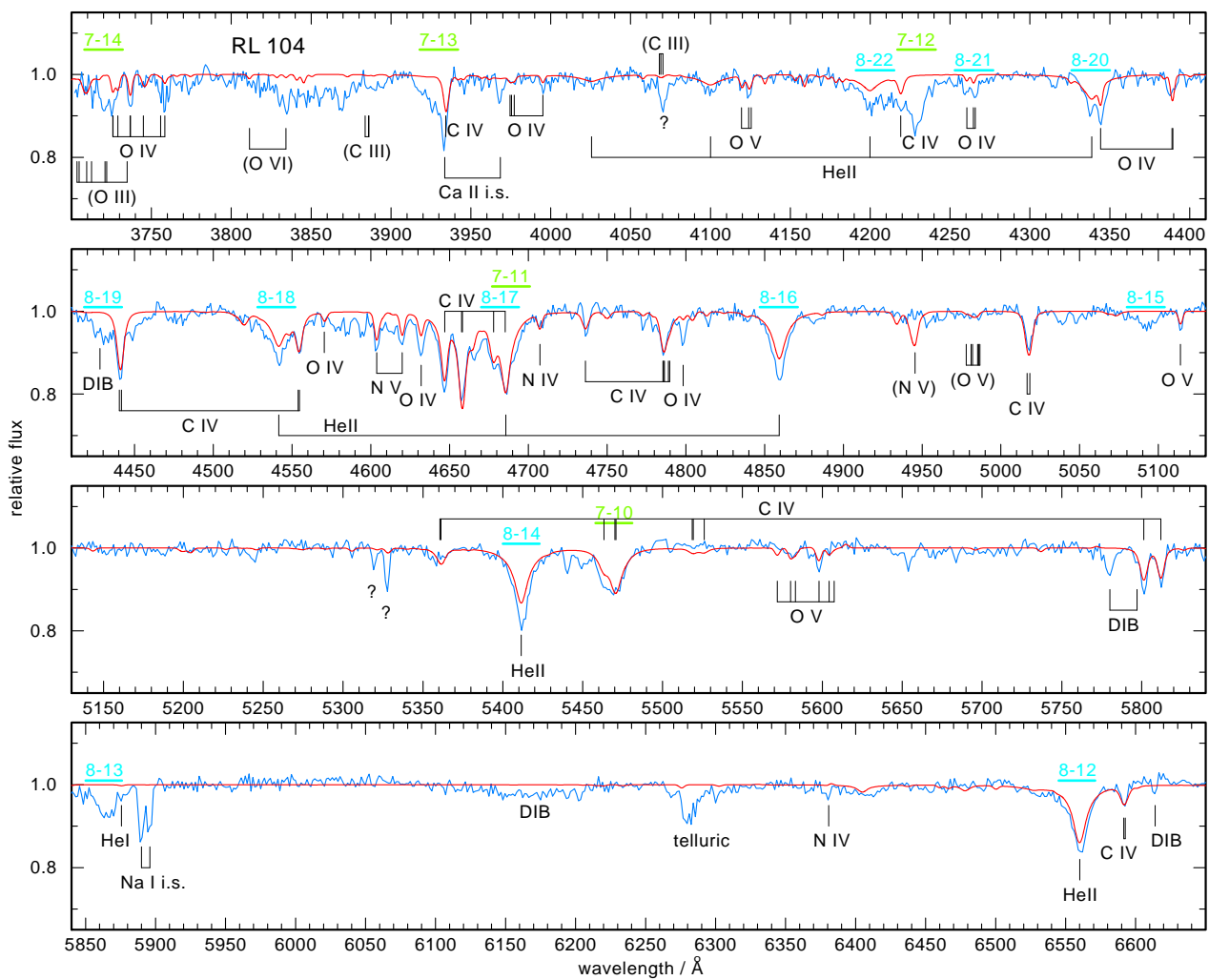}
  \caption{LAMOST spectrum of the PG1159 star \rl (blue
    graph). Overplotted is the final model (red) with \Teff =
    80\,000\,K, \logg = 6.0, and element abundances as given in
    Table\,\ref{tab:resultsall}. We note the unusual presence of highly
    excited \ion{C}{iv} lines at locations indicated by horizontal
    blue and green bars (with principal quantum numbers marked). They
    are not included in the model. Line identifications in brackets
    denote lines that are not observed but potentially constrain the
    atmospheric parameters as discussed in the text. Diffuse
      interstellar bands (DIBs) are indicated.}
\label{fig:rl104_opt}
\end{figure*}

\section{Introduction}
\label{sect:intro}

Stars that are about to become a white dwarf (WD) and that appear
hydrogen deficient defy canonical single-star evolution, because it
predicts that hydrogen-rich envelopes are retained at all
times. Therefore, the phenomenon of hydrogen-deficiency occurring in
objects being in post-asymptotic giant branch (AGB) and post-red giant
branch (RGB) stages is lively debated. Only a small fraction of
pre-WDs are affected. In particular, hot compact hydrogen-deficient
pre-white dwarfs (pre-WDs) with effective temperature \Teff $>$
70\,000\,K and surface gravity 5.0 $>$ \logg $>$ 7.0 are quite scarce.
It is believed that they are the
outcome of either single star evolution (late helium-shell flash or
late helium-core flash) or binary star evolution (double white dwarf
merger). Observationally, such objects are members of three different
spectroscopic classes, namely PG1159, O(He), and He-sdO.

The PG1159 class comprises about 50 WDs and pre-WDs with He-C-O
dominated atmospheres exhibiting large amounts of carbon and oxygen
(typically, He=0.33, C=0.50, O=0.17, mass fractions) with \Teff =
75\,000--250\,000\,K and \logg = 5.3--8.0
\citep[e.g.][]{2006PASP..118..183W,2016A&A...587A.101R}. They are
evolving along post-AGB tracks and are thought to result from a late
He-shell flash.  About one third of them have $\logg<7$ and can be
considered as He-shell burning pre-WDs. Their temperatures are in the \Teff = 100\,000--170\,000\,K range.

O(He) stars form a small group of ten members in the \Teff =
80\,000--200\,000\,K range, \logg = 5.0--6.7, and they have helium-dominated
atmospheres with much less C and O content than PG1159 stars,
typically below 1\% \citep{2014A&A...566A.116R, 2016A&A...587A.101R,
  2014A&A...564A..53W, 2015MNRAS.448.3587D,
  2015A&A...583A.131W}. Although they occupy the same region in the
\logg\--\Teff diagram as the PG1159 stars, the majority of them,
that is, those not surrounded by a planetary nebula, are
believed to be the merger product of two WDs. They are He-shell
burners and probably evolved versions of the cooler R Coronae Borealis
(R\,CrB) stars and extreme helium (EHe) stars
\citep{2002MNRAS.333..121S}, moving towards the hot end of the WD
cooling sequence. The evolutionary status of the three O(He) stars,
which are central stars of planetary nebulae, is less clear.  These
stars must have formed in a different way, because the timescales of
the double WD merger scenario cannot account for their H-rich nebulae
\citep{2014A&A...566A.116R, 2014MNRAS.440.1345F}. 

As with O(He) stars, the helium-rich subdwarf O stars (He-sdOs) have
helium-dominated atmospheres but usually cluster around \Teff =
40\,000--50\,000\,K and \logg = 5.5--6.0 \citep{2016PASP..128h2001H};
hence, they are less luminous than post-AGB stars in that temperature
range. They are either on or above the helium main sequence, they are thus He-core or He-shell burners, respectively. As to their
origin, one favours two competing scenarios, namely a late He-core
flash in a single star or a binary He-WD merger.

A total of 244 He-sdOs with measured \Teff\ and \logg are listed in
the catalogue of hot subdwarfs by \citet{2020A&A...635A.193G}. Only a few
of them are rather hot. Just 13 (about 5\%) have temperatures of
70\,000\,K and higher, and the hottest is about 80\,000\,K. These are
particularly interesting because they are located well below the
helium main sequence, and thus their evolutionary state is
unclear. They could be evolved versions of the common He-sdOs
proceeding onwards to the hot end of the WD cooling sequence. 

In summary, only about three dozen hot H-deficient pre-WDs (\Teff $>$
70\,000\,K, 5.0 $>$ \logg $>$ 7.0) are known and only a fraction of
them have been analysed in detail concerning element abundances. Here,
we present the spectroscopic analysis of five such stars
(Table~\ref{tab:stars}) with parameters in the \Teff
= 70\,000--80\,000\,K range and \logg = 5.2--6.3. One is a new PG1159 star
(\rl) and another one is a reclassified O(He) star (\jl). Then, we have
three He-sdOs (\sdss, \gaiaf, and \gaias). The first of them is a star
previously classified as O(He) and the other two are new He-sdOs. We now briefly introduce our programme stars.

\begin{table*}[t]
\begin{center}
\caption{
Programme stars analysed in this paper. 
  \tablefootmark{a}
}
\label{tab:stars} 
\begin{tabular}{cccccc}
\hline 
\hline 
\noalign{\smallskip}
Full name & Abbreviated name & $\alpha$&$\delta$& V-mag  & Ref. \\ \hline
\noalign{\smallskip} 
\rl    &         & 04 30 14.86 & $+$40 24 14.47 & 13.768 & (1)  \\
\jl    &         & 19 02 32.12 & $-$75 46 34.87 & 13.374 & (2)  \\
\sdss  & \sdsss  & 14 18 12.51 & $-$02 44 27.00 & 16.783 & (1)  \\
\gaiaf & \gaiaff & 11 06 47.78 & $-$57 20 57.00 & 15.071 & (1)  \\
\gaias & \gaiass & 18 45 13.13 & $-$33 05 38.20 & 15.701 & (3)  \\
\noalign{\smallskip}
\hline
\end{tabular} 
\tablefoot{  \tablefoottext{a}{Coordinates from Gaia DR2
    \citep{2018yCat.1345....0G}.  References for V-magnitude (last
    column): (1) \citet{2019A&A...621A..38G}, (2)
    \citet{2020A&A...635A.193G}, (3) \citet{2019AJ....158..138S}. }}
\end{center}
\end{table*}

\paragraph{\rl\ -- PG1159}
was found in a photographic survey as a faint blue star close to
the Galactic plane in the direction of the anticentre
\citep{1974AJ.....79.1406R}. \citet{1979AJ.....84..534C} classified
it as a peculiar DA white dwarf. We discovered that it is a new
PG1159 star. With \Teff = 80\,000\,K and \logg = 6.0, it is located
amongst the hottest He-sdOs; hence, it is less luminous than stars on
post-AGB evolutionary tracks, and therefore is evidence that the PG1159
class of stars might also be fed by another evolutionary channel
besides the late helium-shell flash, namely a binary WD
merger. Because of its relatively low temperature and gravity, \rl
exhibits an unusual optical spectrum as displayed in
Fig.\,\ref{fig:rl104_opt}. It is garnished with absorption lines of
\ion{C}{iv} involving highly excited Rydberg states with quantum
numbers as large as $n=22$.

\begin{figure}[t]
 \centering  \includegraphics[width=0.9\columnwidth]{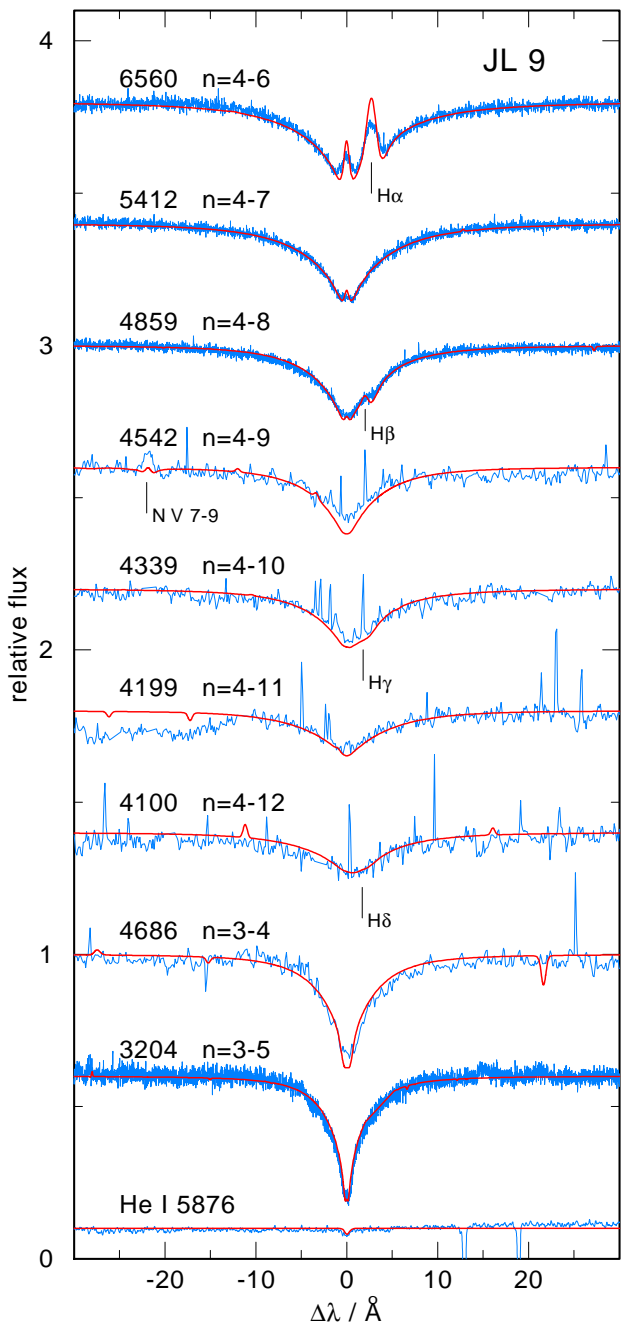}
  \caption{\ion{He}{ii} lines in CASPEC and UVES spectra of the
    O(He) star \jl (blue). At the bottom, a very weak \ion{He}{i} line
    is depicted. The many spikes in the CASPEC spectra are caused by
    cosmic ray hits. Overplotted in red is our final model with \Teff
    = 80\,000\,K, \logg = 5.2, and element abundances as given in
    Table\,\ref{tab:resultsall}.}
\label{fig:jl9_he}
\end{figure}

\begin{figure*}[t]
 \centering  \includegraphics[width=0.9\textwidth]{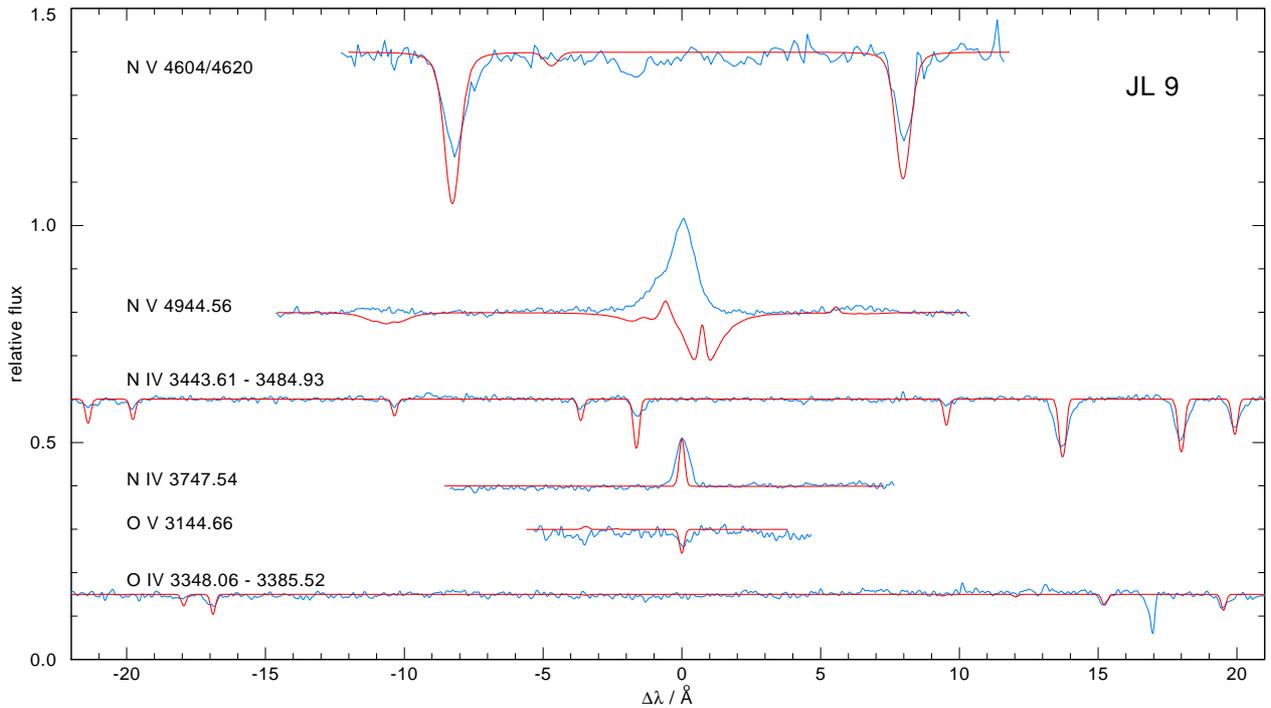}
  \caption{Lines from \ion{N}{iv-v} and \ion{O}{iv-v} in CASPEC and UVES 
    spectra of the O(He) star \jl. Overplotted in red is our final
    model with \Teff = 80\,000\,K, \logg = 5.2, and element abundances
    as given in Table\,\ref{tab:resultsall}.}
\label{fig:jl9_opt}
\end{figure*}

\begin{figure*}[t]
 \centering
 \includegraphics[width=0.28\textwidth,angle=270]{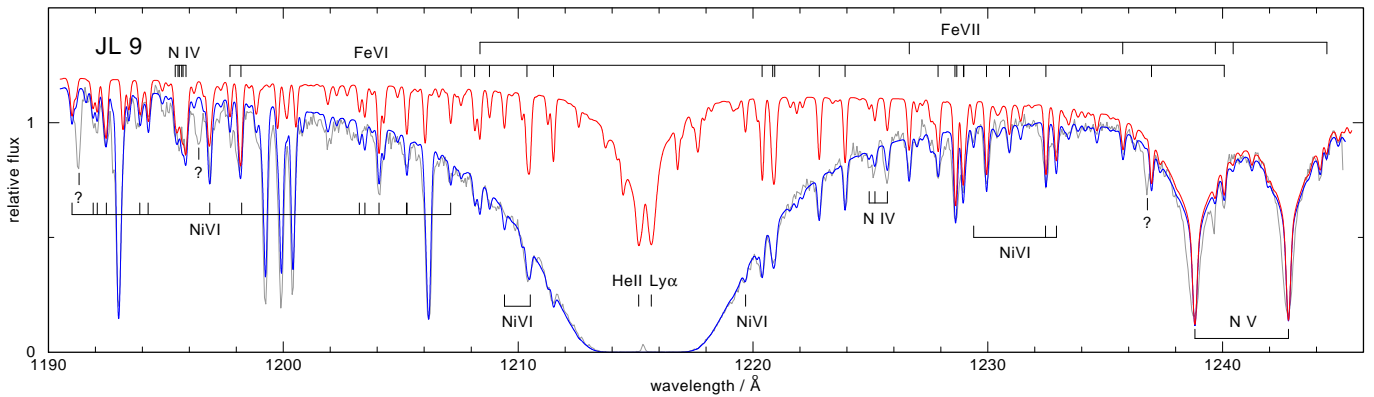}
  \caption{HST/STIS spectrum of the O(He) star \jl\ compared
    to the final model spectrum (red graph; \Teff = 80\,000\,K,
    \logg = 5.2) with the element abundances given in 
    Table\,\ref{tab:resultsall}. The same model attenuated by interstellar lines is
    plotted in blue. Prominent photospheric lines are
    identified. Question marks indicate unidentified 
    lines. A reddening of \ebv = 0.06 was applied to the model spectra.}
\label{fig:jl9_hst}
\end{figure*}

\begin{figure*}[t]
 \centering  \includegraphics[width=1.0\textwidth]{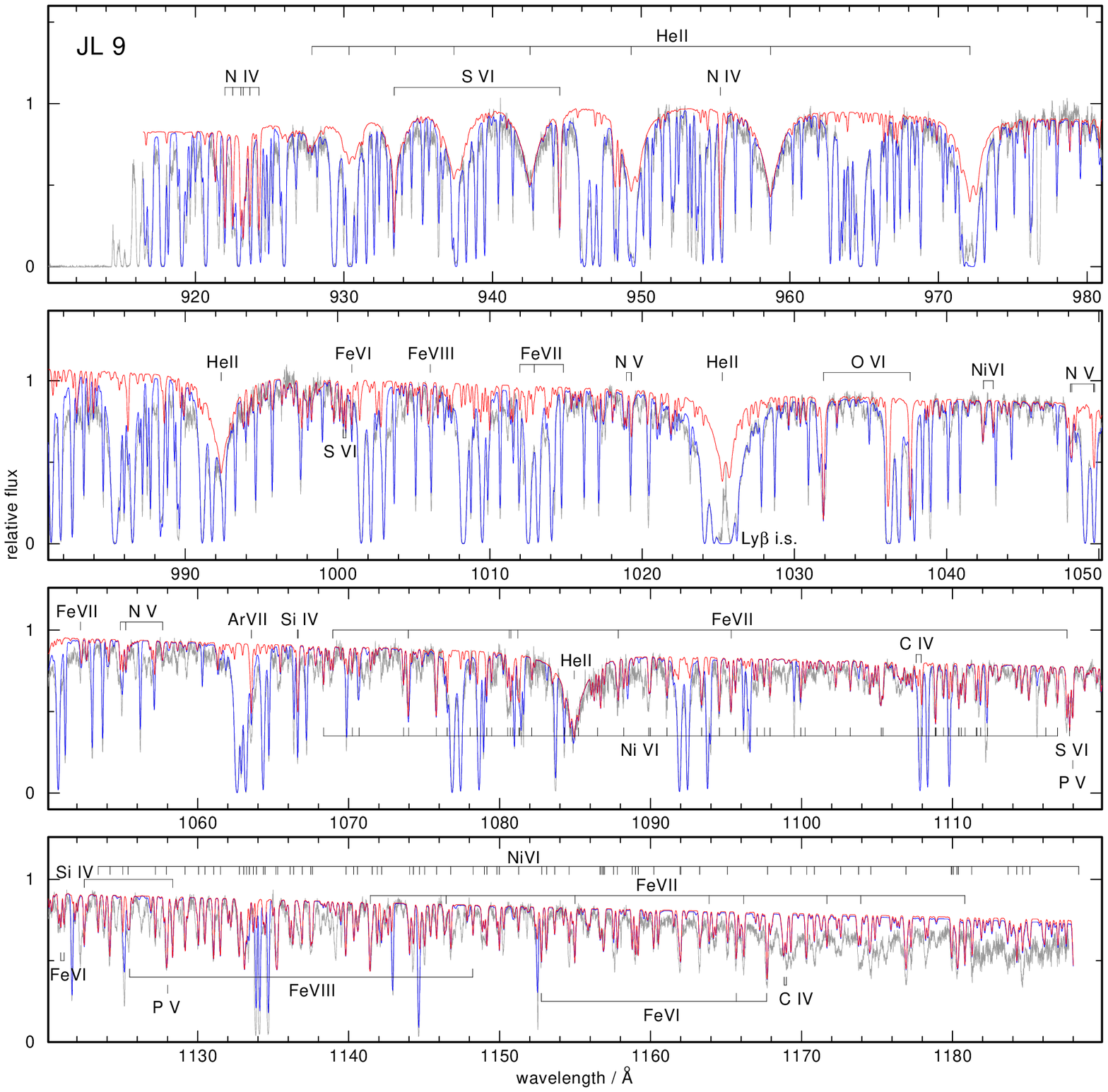}
  \caption{FUSE spectrum of the O(He) star \jl\ compared
    to the final model spectrum (red graph; \Teff = 80\,000\,K,
    \logg = 5.2) with the element abundances given in 
    Table\,\ref{tab:resultsall}. The same model attenuated by interstellar lines is
    plotted in blue.}
\label{fig:jl9_fuse}
\end{figure*}

\paragraph{\jl\  -- O(He)}
 first appeared in a catalogue of faint ultraviolet stars from a
photographic survey \citep{1969ArA.....5..345J}. It was classified as
He-sdO by \cite{1986ASSL..128..345H}, and its temperature was estimated
at \Teff = 80\,000\,K from the UV flux distribution measured with the
International Ultraviolet Explorer (IUE). No further analyses have been
published since then. We state here that the star must be
re-classified and is in fact of spectral type O(He) (\Teff =
80\,000\,K, \logg = 5.2), meaning it is rather luminous compared to usual
He-sdOs. Together with another O(He) star, it is the coolest member of
this class and thus represents a link to EHe stars. We analysed
archival optical and ultraviolet (UV) spectra
(Figs.\,\ref{fig:jl9_he}--\ref{fig:jl9_nv4945}).

\paragraph{\sdss\ -- He-sdO}
 has been classified as an O(He) star based on an analysis of
SDSS spectra revealing \Teff = $90\,000 \pm 20\,000$\,K and \logg =
$5.5\pm0.5$ \citep{2014A&A...564A..53W}. With the analysis of new
high-quality optical spectra (Fig.\,\ref{fig:sdssj141812_opt}), we show
here that the atmospheric parameters are at the cool and high-gravity
end of the error margins of the previous analysis (namely \Teff =
$70\,000$\,K and \logg = $6.0$). Consequently, we reclassify the star as
He-sdO. We also find the presence of trace hydrogen in the new
spectra. For conciseness, we abbreviate the name of this
star as \sdsss\ hereafter.

\begin{table}[t]
\begin{center}
\caption{
Metal lines identified in the CASPEC and UVES spectra of \jl. 
  \tablefootmark{a}
}
\label{tab:lines} 
\begin{tabular}{llcl}
\hline 
\hline 
\noalign{\smallskip}
Wavelength / \AA & Ion & Transition &\\ \hline
\noalign{\smallskip} 
3443.61 & \ion{N}{iv} & $2{\rm p}\ 3{\rm s}\ ^3{\rm P}^{\rm o}_{1}-2{\rm p}\ 3{\rm p}\ ^3{\rm D}_{2}$ \\
3445.22 & \ion{N}{iv} & $2{\rm p}\ 3{\rm s}\ ^3{\rm P}^{\rm o}_{0}-2{\rm p}\ 3{\rm p}\ ^3{\rm D}_{0}$ \\
3454.65 & \ion{N}{iv} & $2{\rm p}\ 3{\rm s}\ ^3{\rm P}^{\rm o}_{1}-2{\rm p}\ 3{\rm p}\ ^3{\rm D}_{0}$ \\ 
3461.36 & \ion{N}{iv} & $2{\rm p}\ 3{\rm s}\ ^3{\rm P}^{\rm o}_{1}-2{\rm p}\ 3{\rm p}\ ^3{\rm D}_{0}$ \\ 
3463.36 & \ion{N}{iv} & $2{\rm p}\ 3{\rm s}\ ^3{\rm P}^{\rm o}_{2}-2{\rm p}\ 3{\rm p}\ ^3{\rm D}_{2}$ \\ 
3474.53 & \ion{N}{iv} & $2{\rm p}\ 3{\rm s}\ ^3{\rm P}^{\rm o}_{2}-2{\rm p}\ 3{\rm p}\ ^3{\rm D}_{2}$ \\
3478.72 & \ion{N}{iv} & $2{\rm s}\ 3{\rm s}\ ^3{\rm S}_{1}-2{\rm p}\ 3{\rm p}\ ^3{\rm P}^{\rm o}_{2}$ \\ 
3483.00 & \ion{N}{iv} & $2{\rm s}\ 3{\rm s}\ ^3{\rm S}_{1}-2{\rm p}\ 3{\rm p}\ ^3{\rm P}^{\rm o}_{1}$ \\  
3484.93 & \ion{N}{iv} & $2{\rm s}\ 3{\rm s}\ ^3{\rm S}_{1}-2{\rm p}\ 3{\rm p}\ ^3{\rm P}^{\rm o}_{0}$ \\   
3747.54em &\ion{N}{iv} & $2{\rm p}\ 3{\rm s}\ ^1{\rm P}^{\rm o}_{1}-2{\rm p}\ 3{\rm p}\ ^1{\rm D}_{2}$ \\
4057.76em &\ion{N}{iv} & $3{\rm p}\ ^1{\rm P}^{\rm o}_{1}-3{\rm d}\ ^1{\rm D}_{2}$ \\
5200.41em& \ion{N}{iv}& $2{\rm p}\ 3{\rm s}\ ^3{\rm P}^{\rm o}_{1}-2{\rm p}\ 3{\rm p}\ ^3{\rm D}_{2}$ \\
5204.28em& \ion{N}{iv}& $2{\rm p}\ 3{\rm s}\ ^3{\rm P}^{\rm o}_{2}-2{\rm p}\ 3{\rm p}\ ^3{\rm D}_{3}$ \\
\noalign{\smallskip}
3159.76 & \ion{N}{v}  &  5p $-$ 6s \\
3161.37 & \ion{N}{v}  &  5p $-$ 6s \\
3501.22em & \ion{N}{v}  &  7i $-$ 10k etc.\\
4519.87em & \ion{N}{v}  &  7i $-$ 9k etc.\\
4603.74 & \ion{N}{v}  &  3s $-$ 3p\\
4619.97 & \ion{N}{v}  &  3s $-$ 3p\\
4934.16em& \ion{N}{v}  &  6d $-$ 7f \\
4944.56em & \ion{N}{v}  &  6f $-$ 7g, 6g $-$ 7h,  6h $-$ 7i  \\
6478.50em & \ion{N}{v}  &  8k $-$ 10l etc.\\
\noalign{\smallskip}
3063.43 & \ion{O}{iv} & $3{\rm s}\ ^2{\rm S}_{1/2}-3{\rm p}\ ^2{\rm P}^{\rm o}_{3/2}$ \\
3071.60:& \ion{O}{iv} & $3{\rm s}\ ^2{\rm S}_{1/2}-3{\rm p}\ ^2{\rm P}^{\rm o}_{1/2}$ \\ 
3348.06:& \ion{O}{iv} & $3{\rm s}\ ^2{\rm P}^{\rm o}_{1/2}-3{\rm p}\ ^2{\rm D}_{3/2}$ \\
3349.11 & \ion{O}{iv} & $3{\rm s}\ ^2{\rm P}^{\rm o}_{3/2}-3{\rm p}\ ^2{\rm D}_{5/2}$ \\
3381.21 & \ion{O}{iv} & $3{\rm s}\ ^4{\rm P}^{\rm o}_{3/2}-3{\rm p}\ ^4{\rm D}_{5/2}$ \\
3381.30 & \ion{O}{iv} & $3{\rm s}\ ^4{\rm P}^{\rm o}_{1/2}-3{\rm p}\ ^4{\rm D}_{3/2}$ \\ 
3385.52 & \ion{O}{iv} & $3{\rm s}\ ^4{\rm P}^{\rm o}_{5/2}-3{\rm p}\ ^4{\rm D}_{7/2}$ \\ 
\noalign{\smallskip} 
3144.66 & \ion{O}{v}  & $2{\rm s}\ 3{\rm p}\ ^1{\rm P}^{\rm o}_{1}-2{\rm s}\ 3{\rm d}\ ^1{\rm D}_{2}$ \\
\noalign{\smallskip}
\hline
\end{tabular} 
\tablefoot{  \tablefoottext{a}{Colons indicate very weak lines with
    uncertain identifications and 'em' emission lines.  The
    \ion{N}{v} line at 4944.56\,\AA\ is an unresolved emission line
    blend (Fig.\,\ref{fig:jl9_nv4945} and Table~\ref{tab:nv}); see
    text for details.}  } 
\end{center}
\end{table}

\begin{figure*}[t]
 \centering  \includegraphics[width=0.9\textwidth]{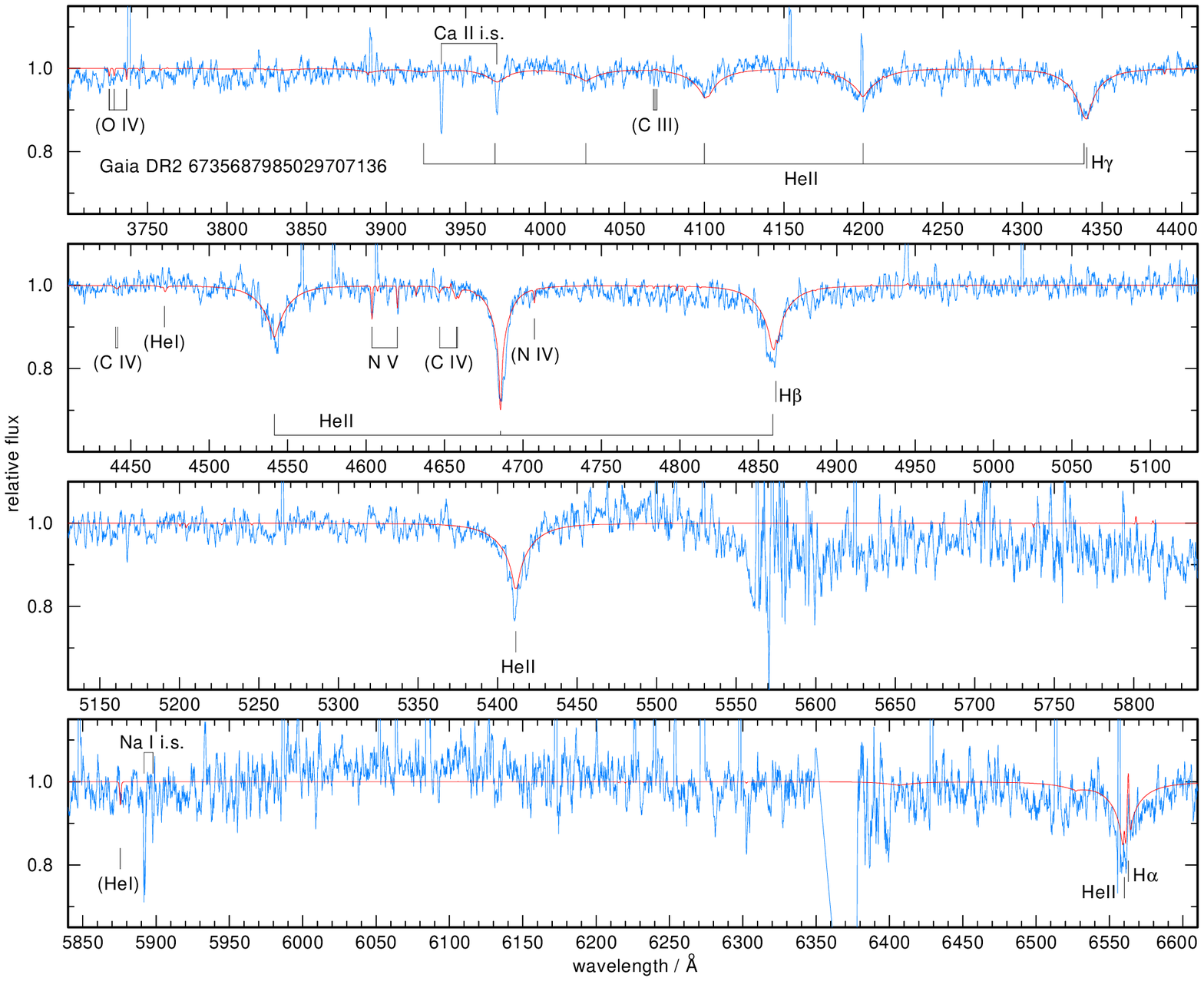}
  \caption{X-shooter spectrum of the He-sdO \gaiass. Emission spikes (except H$\alpha$) are
    cosmic ray hits. Overplotted is the
    final model (red) with \Teff = 75\,000\,K, \logg = 6.3, and element
    abundances as given in Table\,\ref{tab:resultsall}. }
\label{fig:gaias_opt}
\end{figure*}

\paragraph{\gaiaf and \gaias\ -- He-sdOs}
 are two new He-sdOs. We analysed optical spectra
(Figs.\,\ref{fig:gaiaf_opt} and \ref{fig:gaias_opt}) and find that
they have similar atmospheric parameters. Both have \Teff =
$70\,000$\,K and  \logg = $6.0$ and $6.3$, respectively, plus trace
hydrogen. Hence, within our sample they, together with \sdsss, form a
triplet of He-sdOs with similar temperatures and gravities. In the
following, we abbreviate the names of these stars as \gaiaff and
\gaiass.

In Section\,\ref{sect:observations}, we present the
observations. We then turn to the spectral analysis of the stars
(Sect.\,\ref{sect:analysis}) and derive their physical parameters
in Sect.\,\ref{sect:sed}. The results are summarized and
discussed in Sect.\,\ref{sect:discussion}.

\section{Observations}
\label{sect:observations}

\subsection{\rl\ -- Optical spectra}

The star caught our attention from a spectrum taken on December 13,
2020, with the Asiago 1.22\,m telescope and the B\&C spectrograph.
The star was also observed with LAMOST, DR6
\citep{2020ApJ...889..117L} and this spectrum is analysed in the
present paper. It covers the 3700--9080\,\AA\ range with a resolving
power of $R \approx 1500$. It exhibits features characteristic for
relatively `cool' (\Teff $\approx 80\,000$\,K) PG1159 stars
\citep{2014A&A...569A..99W}. It is dominated by \ion{C}{iv} and
\ion{He}{ii} lines, together with lines of \ion{O}{iv} and \ion{O}{v}
(Fig.\,\ref{fig:rl104_opt}). We also spot \ion{N}{iv} and \ion{N}{v}
lines. What is significantly different from all other `cool' PG1159
stars are the deep and narrow \ion{C}{iv} and \ion{He}{ii} lines,
immediately pointing at a much lower gravity than the usual value of
\logg $\approx 7$, indicating that the star is not in the WD range of
the Hertzsprung-Russell diagram (HRD), but that it is a luminous
pre-WD. As a consequence of the low gravity and hence lower
atmospheric particle densities, rather unusual absorption features
appear, which we identify as \ion{C}{iv} lines involving extremely
high excited energy levels. While the \ion{C}{iv} lines commonly
observed in PG1159 stars stem from levels with principal quantum
numbers $n\leq10$ (e.g. $4-5$ and $6-8$ transitions in the absorption
trough around \ion{He}{ii}~4686, or $7-10$ at 5470\,\AA), we see in
\rl\ broad absorption lines that stem from transitions starting from
levels with $n=7$ and 8 going into levels with $n'= 11$ and up to $n'=
22$. We note that photospheric lines from other elements in H-like ionisation stages
(\ion{O}{vi}, \ion{Ne}{viii}) involving levels with high principal
quantum numbers were identified in PG1159 stars, [WCE]
central stars, and a DO white dwarf \citep{2007A&A...474..591W}.
In accordance with the high reddening of the star
(Sect.\,\ref{sect:sed}), the spectrum
exhibits diffuse interstellar bands \citep[see e.g. line lists
  in][]{2008ApJ...680.1256H}.

\subsection{\jl\ -- Optical and ultraviolet spectra}

To probe the interstellar medium, \jl has been observed at UV
wavelength with IUE, FUSE, HST, and the ESO VLT \citep[e.g.
][]{1994ApJS...93..211D,2005ApJ...635.1136H,2008A&A...481..381L,2010MNRAS.404.1321W,2013ApJ...764...25J}.
These spectra contain a lot of information on the photosphere of \jl,
which has not been exploited yet.  We retrieved an archival spectrum
of \jl\ that was recorded on May 19, 2006, with the UVES spectrograph
at the ESO Very Large Telescope (ProgID: 077.C-0547(A); PI:
R. Lallement). It covers the wavelength regions 3050--3870 and
4785--6809\,\AA\ with a resolving power of $R \approx 40\,000$.
Because the UVES observations were tailored to probe the ISM, the
wavelength range between $\approx$\,4000 and 5000\AA\ was not
covered. However, many important diagnostic lines for the photospheric
analysis are located in this range. Therefore, we complemented the
UVES spectrum with a spectrum of lower resolution to fill the gap.  We
employed a spectrum taken on October 6, 1984, with the CASPEC
spectrograph attached to the ESO 3.6m telescope
\citep[see][]{1986ASSL..128..345H,1987fbs..conf..603D}. It covers
4042--5085\,\AA\ with a spectral resolution of 0.5\,\AA. It suffered
from cosmic ray hits that were not removed. 

The spectra are dominated by \ion{He}{ii} lines, and one very weak
neutral helium line (5876\,\AA) can be detected
(Fig.\,\ref{fig:jl9_he}). Contributions of weak hydrogen Balmer lines
are visible in the respective \ion{He}{ii} line profiles. Both the
H$\alpha$ and the adjacent \ion{He}{ii} line cores are in emission.
About thirty metal lines can be identified (Table\,\ref{tab:lines},
Fig.\,\ref{fig:jl9_opt}) and some of them are in emission. They stem
from \ion{N}{iv-v} and \ion{O}{iv-v}. Other metals, in particular
carbon, cannot be detected.

We utilised an archival UV spectrum taken with the STIS instrument
aboard the Hubble Space Telescope (HST, dataset OBIE16010). It was
taken on July 22, 2011, with grating G140M, and it covers the 1191--1245.5\,\AA\ wavelength
range with a resolution of
0.1\,\AA\ (Fig.\,\ref{fig:jl9_hst}). We identified lines from nitrogen,
iron, and nickel; namely, \ion{N}{iv}, \ion{N}{v}, \ion{Fe}{vi},
\ion{Fe}{vii}, and \ion{Ni}{vi}.

In addition, we used two archival far-UV observations taken by the Far
Ultraviolet Spectroscopic Explorer (FUSE, data IDs P302120100,
U10946010031) with the LWRS aperture that cover the
905--1188\,\AA\ range with $R \approx 15\,000$. These observations
consist of five exposures in total with a combined exposure time of
about 21\,000\,s. We retrieved the calibrated and extracted spectra
from the Mikulski Archive for Space Telescopes (MAST) for each of the
eight detector segment and channel combinations. For each combination,
the five exposures were cross-correlated and then co-added. The final
spectrum is affected by the `worm', a grid wire shadow on the
detector, causing a flux depression longwards of about 1160\,\AA.  We
identify broad lines from \ion{He}{ii}, every other one being blended by
interstellar hydrogen Lyman lines (Fig.\,\ref{fig:jl9_fuse}). The
majority of lines stems from \ion{Ni}{vi}, but we also see lines from
iron and from several light metals. They are discussed in detail
below.

The far-UV spectrum of \jl is affected by strong interstellar lines
and was used by
\cite{2004ApJ...609..838W} and \cite{2013ApJ...764...25J} to study the interstellar medium (ISM). To account
for blends between photospheric and interstellar lines, we constructed
a simple model for the ISM. This model uses the interstellar line list
distributed with the ISM code OWENS \citep{2002ApJS..140...67L} and
considers several IS clouds that are allowed to vary in radial
velocity, Doppler broadening, and composition. Our model for \jl
includes observed lines of \ion{H}{i}, \ion{D}{i}, \ion{C}{i-ii}, \ion{N}{i-ii},
\ion{O}{i}, \ion{O}{vi},  \ion{Si}{ii-iii}, \ion{P}{ii},
\ion{Ar}{i}, \ion{Mn}{ii}, \ion{Fe}{i-ii}, as well as H$_2$ and HD. We
refer the reader to \cite{2004ApJ...609..838W} for a quantitative
description of the ISM towards \jl.

\begin{figure}[t]
 \centering  \includegraphics[width=0.9\columnwidth]{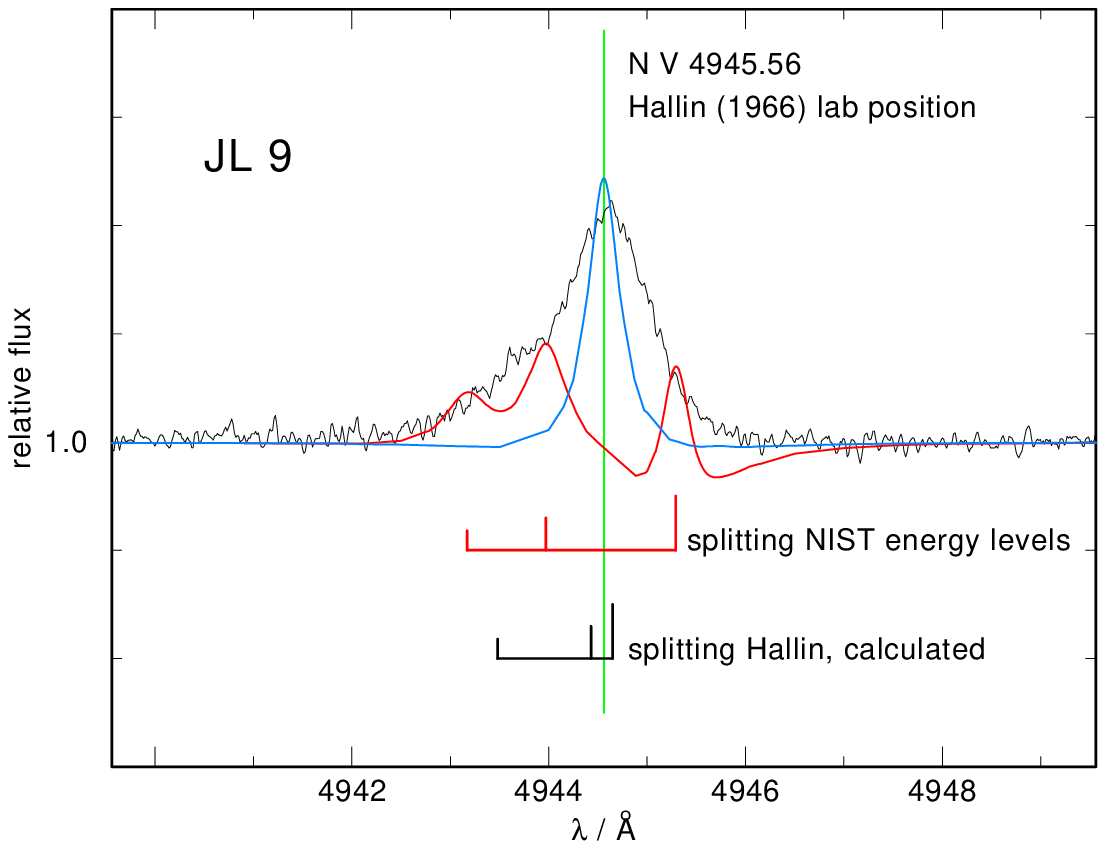}
  \caption{ \ion{N}{v} 4946\,\AA\ emission line in the UVES spectrum of \jl. It is a
    blend of three components with uncertain wavelength splittings. Their
    positions from NIST energy levels and from \cite{Hallin1966}
    are indicated, where the length of the vertical bars scales
    with the gf-value. Red graph: Model assuming NIST splitting. Blue
    graph: Model assuming zero splitting.}
\label{fig:jl9_nv4945}
\end{figure}

\begin{figure*}[t]
 \centering  \includegraphics[width=0.9\textwidth]{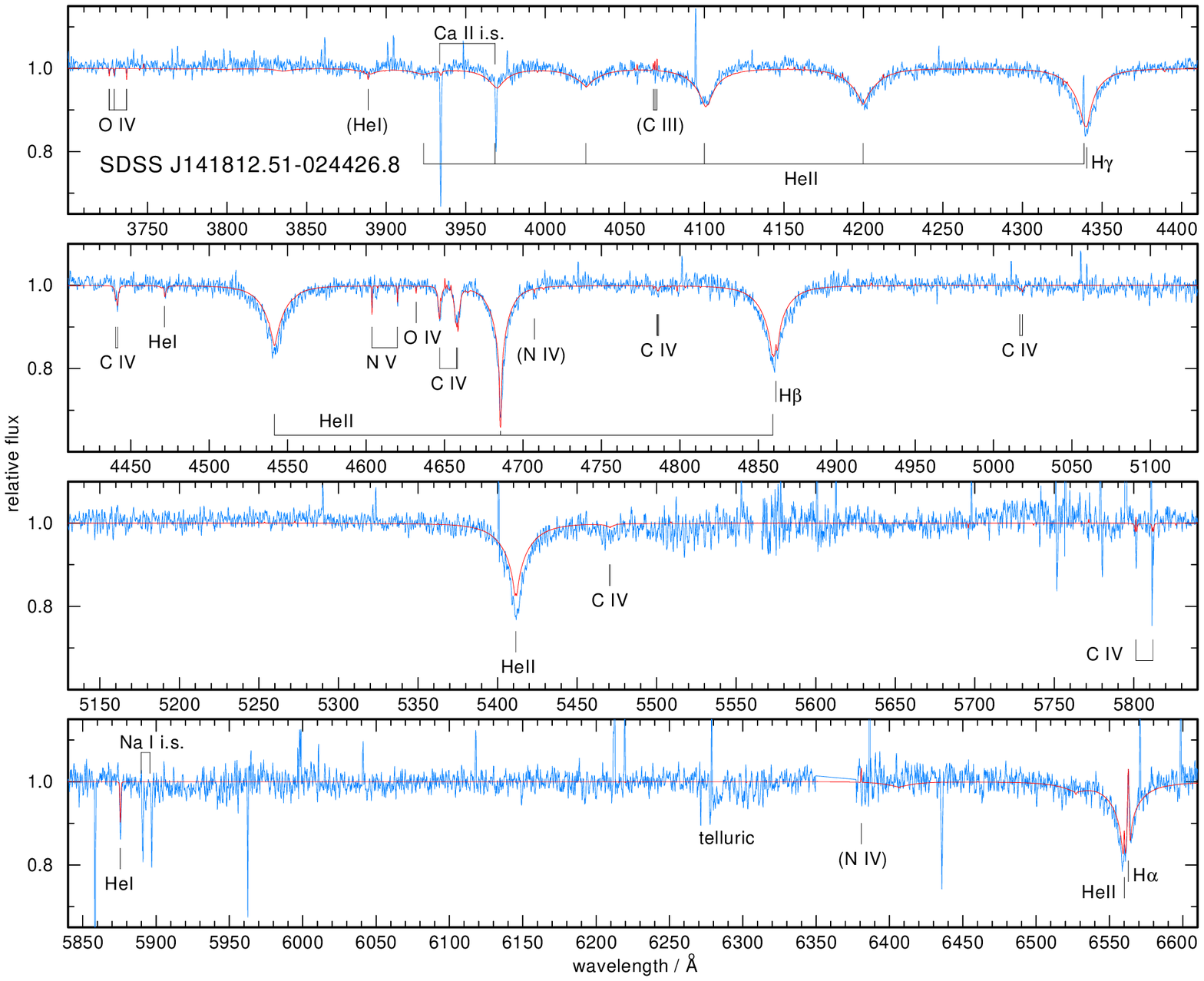}
  \caption{X-shooter spectrum of the He-sdO \sdss. Emission spikes (except H$\alpha$) are
    cosmic ray hits. Overplotted is the final model (red) with \Teff =
    70\,000\,K, \logg = 6.0, and element abundances as given in
    Table\,\ref{tab:resultsall}. }
\label{fig:sdssj141812_opt}
\end{figure*}

\subsection{\sdsss, \gaiaff, and \gaiass\ -- Optical spectra}

Optical spectra were obtained with the X-shooter instrument at ESO's
Very Large Telescope on May 16, 2016 (ProgID: 297.D-5004(A); PI:
S. Geier). They cover the wavelength range 3000--10\,000\,\AA\ with $R
\approx 10\,000$.  With the same instrument setup, \gaiaff and \gaiass
were observed on April 10, 2021 (ProgID: 105.206H.001; PI:
S. Geier). The spectra of all three He-sdOs are rather similar. They
are dominated by \ion{He}{ii} lines with weak or absent \ion{He}{i}
lines, indicating a very high temperature at first glance. They all
show an H$\alpha$ line emission core, revealing trace amounts of
hydrogen. They also exhibit prominent \ion{C}{iv} lines, with the
conspicuous exception of \gaiass and \ion{N}{v} lines.

\begin{figure*}[t]
 \centering  \includegraphics[width=0.9\textwidth]{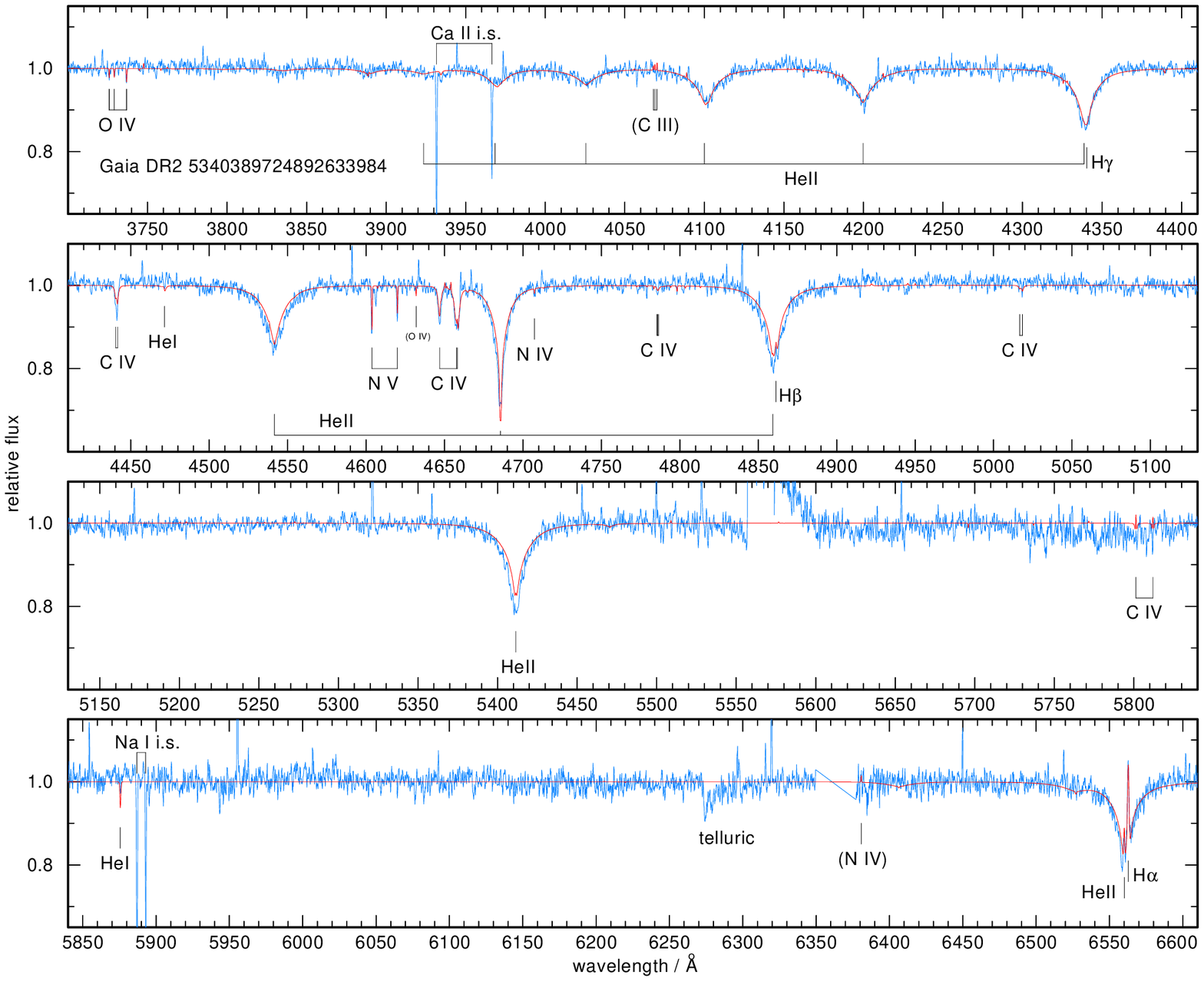}
  \caption{X-shooter spectrum of the He-sdO \gaiaff. Emission spikes (except H$\alpha$) are
    cosmic ray hits. Overplotted is the
    final model (red) with \Teff = 75\,000\,K, \logg = 6.0, and element
    abundances as given in Table\,\ref{tab:resultsall}. }
\label{fig:gaiaf_opt}
\end{figure*}

\begin{table}[t]
\begin{center}
\caption{Components of \ion{N}{v} 4946\,\AA\ emission line blend in \jl.
\tablefootmark{a} }
\label{tab:nv} 
\begin{tabular}{lllll}
\hline 
\hline 
\noalign{\smallskip}
\ion{N}{v}&NIST&Hallin&Rieger et al.&$\log gf$\\
\hline 
\noalign{\smallskip}
$6f-7g$ & 4943.17 & 4943.48 & 4943.38$\pm$0.40 & 1.19 \\
$6g-7h$ & 4943.97 & 4944.43 & 4944.40$\pm$0.43 & 1.43 \\
$6h-7i$ & 4945.29 & 4944.65 & 4944.40$\pm$0.43 & 1.63 \\
\hline
\end{tabular} 
\tablefoot{  \tablefoottext{a}{See also
    Fig.\,\ref{fig:jl9_nv4945}. Wavelength positions in \AA. The column `NIST'
    lists positions computed from level energies at NIST. References:
    \cite{Hallin1966}, \cite{1995PhyS...52..166R}.}  } 
\end{center}
\end{table}

\begin{table*}[t]
\begin{center}
\caption{Atmospheric properties and derived stellar parameters of the analysed stars.
\tablefootmark{a} }
\label{tab:resultsall}
\begin{tabular}{rrrrrrr}
\hline 
\hline 
\noalign{\smallskip}
                          & \rl                & \jl                & \sdsss                & \gaiaff             & \gaiass &           Sun \\ 
\hline 
\noalign{\smallskip}
spectral type             & PG1159             & O(He)              & He-sdO                & He-sdO             & He-sdO                &\\
\Teff/\,K                 & $80\,000 \pm 5000$ & $80\,000 \pm 5000$ & $70\,000 \pm 5000$    & $75\,000 \pm 5000$ & $75\,000 \pm 5000$    &\\
$\log$($g$\,/\,cm\,s$^{-2}$) & $6.0 \pm 0.5$     & $5.2 \pm 0.3$    & $6.0 \pm 0.3$         & $6.0 \pm 0.3$      & $6.3 \pm 0.3$          &\\
\noalign{\smallskip}
H                         &$<3.0 \times 10^{-3}$&$0.058^{+0.029}_{-0.019}$& $0.10\pm0.04$     & $0.10\pm0.04$      & $0.10\pm0.04$          & 0.74 \\             
He                        & 0.43               & $0.93^{+0.03}_{-0.03}$& $0.90\pm0.04$       & $0.90\pm0.04$      & $0.90\pm0.04$          & 0.25 \\   
\noalign{\smallskip}
C                         & 0.38               & $7.0 \times 10^{-5}$ & $5.7 \times 10^{-3}$ & $5.7 \times 10^{-3}$ & $<1.0 \times 10^{-3}$& $2.4 \times 10^{-3}$\\ 
N                         & 0.02               & $6.9 \times 10^{-3}$ & $5.8 \times 10^{-4}$ & $5.8 \times 10^{-4}$ & $3.5 \times 10^{-3}$ & $6.9 \times 10^{-4}$\\ 
O                         & 0.17               & $7.9 \times 10^{-5}$ & $6.6 \times 10^{-4}$ & $6.6 \times 10^{-4}$ & $<2.0 \times 10^{-3}$& $5.7 \times 10^{-3}$\\ 
Si                        &                    & $7.9 \times 10^{-4}$ &                     &                      &                     & $6.6 \times 10^{-4}$\\ 
P                         &                    & $2.3 \times 10^{-6}$ &                     &                      &                     & $5.8 \times 10^{-6}$\\ 
S                         &                    & $1.9 \times 10^{-4}$ &                     &                      &                     & $3.1 \times 10^{-4}$ \\ 
Ar                        &                    & $7.3 \times 10^{-5}$ &                     &                      &                     & $8.2 \times 10^{-5}$ \\ 
Fe                        &                    & $4.5 \times 10^{-3}$ &                     &                      &                     & $1.3 \times 10^{-3}$ \\ 
Ni                        &                    & $2.6 \times 10^{-4}$ &                     &                      &                     & $7.1 \times 10^{-5}$ \\ 
\noalign{\smallskip}
\vrad\ / km\,s$^{-1}$      & $+23\pm10$         & $+76.4\pm1.0$        & $-55\pm3$           & $+143\pm3$           & $-71\pm6$ \\
\noalign{\smallskip}
$\log$($L$\,/\,\Lsol)     & $3.13^{+0.15}_{-0.22}$ & $3.36^{+0.15}_{-0.23}$ & $2.34^{+0.22}_{-0.47}$ & $2.46^{+0.17}_{-0.28}$ & $2.81^{+0.24}_{-0.55}$ &\\
\noalign{\smallskip}
$R$\,/\,\Rsol             & $0.19^{+0.02}_{-0.07}$ & $0.25^{+0.04}_{-0.04}$ & $0.10^{+0.03}_{-0.03}$ & $0.10^{+0.02}_{-0.02}$ & $0.15^{+0.05}_{-0.05}$ &\\
\noalign{\smallskip}
$M$\,/\,\Msol\ (VLTP)     & $0.48^{+0.03}_{-0.02}$ & $0.52^{+0.09}_{-0.02}$ \\
\noalign{\smallskip}
$M$\,/\,\Msol\ (merger)   &                    & $0.68^{+0.11}_{-0.06}$ & $0.50^{+0.06}_{-0.06}$ & $0.53^{+0.07}_{-0.05}$ & $0.45^{+0.09}_{-0.05}$ &\\
\noalign{\smallskip}
\ebv\,/ mag               & 0.274              & 0.019                & 0.052               & 0.029               & 0.027                &\\
\noalign{\smallskip}
$H_\nu$ / erg cm$^{-2}$s$^{-1}$Hz$^{-1}$ & $1.307\times 10^{-3}$ & $1.221\times 10^{-3}$ & $1.002\times 10^{-3}$ & $1.073\times 10^{-3}$ & $1.059\times 10^{-3}$\\
\noalign{\smallskip}
$d$ / pc (spectroscopic)  & $683^{+609}_{-307}$  & $2368^{+1273}_{-767}$ &$3359^{+1662}_{-1128}$ & $1681^{+845}_{-548}$  & $1460^{+799}_{-485}$ \\
\noalign{\smallskip}
$d$ / pc (Gaia parallax)  & $1019^{+25}_{-20}$  & $1552^{+70}_{-66}$    &$2819^{+654}_{-423}$  & $1224^{+57}_{-47}$    & $2808^{+577}_{-423}$ \\
\noalign{\smallskip}
\hline
\end{tabular} 
\tablefoot{  \tablefoottext{a}{Element abundances given in mass
    fractions. The error limits for the abundances in \rl are
    $\pm0.3$\,dex. For the metal abundances in the other objects, see
    main text. Solar abundances from
    \citet{2009ARA&A..47..481A}. Luminosities, radii, and reddening
    derived from SED fitting using Gaia parallax distances from
    \cite{2021AJ....161..147B}. Masses derived from VLTP tracks
    (Fig.\,\ref{fig:pg1159_evolution}) or merger tracks
    (Fig.\,\ref{fig:merger_evolution}). $H_\nu$ is the Eddington flux
    of the best-fit atmosphere model used for the spectroscopic
    distance estimate. }  } 
\end{center}
\end{table*}

\section{Spectral analysis}
\label{sect:analysis}

\subsection{The PG1159 star \rl}

We used the T\"ubingen Model-Atmosphere Package
(TMAP\footnote{\url{http://astro.uni-tuebingen.de/~TMAP}}) to compute
non-LTE, plane-parallel, line-blanketed atmosphere models in radiative
and hydrostatic equilibrium
\citep{1999JCoAM.109...65W,2003ASPC..288...31W,tmap2012}. We computed
models of the type introduced in detail by
\cite{2014A&A...569A..99W}. They were tailored to investigate the
optical spectra of relatively cool PG1159 stars. In essence, they
consist of the main atmospheric constituents, namely helium, carbon,
and oxygen. Nitrogen was included as a trace element in subsequent
line-formation iterations, meaning the atmospheric structure was kept
fixed.

A problem was encountered because of the presence of \ion{C}{iv} lines
arising between highly excited levels with principal quantum numbers of
$n=7$ and 8 for the lower levels and $n'=13$ up to 22 for the
upper levels (Fig.\,\ref{fig:rl104_opt}). Due to linear Stark
broadening, they are broad and shallow, and they blend every
\ion{He}{ii} Pickering line (but also the \ion{He}{ii} Fowler $\alpha$
line at 4686\,\AA), which are commonly used to constrain the surface
gravity. For example, \ion{He}{ii}~5411 ($n=4-7$) is blended by
\ion{C}{iv} $n=8-14$. We cannot model these carbon lines because of
the lack of atomic data. However, the effect on the helium line in
this case can be estimated by looking at the next highest member of
the respective \ion{C}{iv} series, namely $n=8-15,$ which is located in
isolation at 5090\,\AA. As can be expected, the \ion{C}{iv} lines with
lower level $n=7$ are even stronger than those with $n=8$, as can be
seen, for example, at the isolated $n=7-12$ line at 4230\,\AA. Probably
the best gravity indicators are the highest members of the
\ion{He}{ii} Pickering series at 4100\,\AA\ and blueward, because the
blending \ion{C}{iv} lines should be much weaker or virtually
absent. With this caveat in mind, we computed a series of model
atmospheres by varying \Teff, \logg\ and element abundances (He, C, N,
O) to gradually approach the final adopted parameter
values. Computing a grid of models by systematically varying all
parameters was prohibitive, because the model atmospheres turned out
to be numerically very unstable in this parameter range. Each single
model required extensive care to bring it to convergence.

Our model fitting procedure arrived at \Teff = $80\,000 \pm 5\,000$\,K
and \logg = $6.0 \pm 0.5$, and abundances He = 0.43, C = 0.38, N =
0.02, O = 0.17 (mass fractions; Table\,\ref{tab:resultsall}). The
uncertainty in the abundances is estimated to be 0.3 dex. An upper
limit of the hydrogen abundance of H $<3.0 \times 10^{-3}$ was
derived. At higher abundances, a H$\alpha$ emission would be visible
in the red wing of the respective \ion{He}{ii} line. The following
considerations constrain the effective temperature and gravity. We can use
the temperature dependence of the relative strengths of the lines from
\ion{N}{iv}/\ion{N}{v} and \ion{O}{iv}/\ion{O}{v}. Looking at models
$\Delta \Teff = \pm 10\,000$\,K, the lines from ionisation stage IV
(e.g. \ion{N}{iv} 4708 and 6380\,\AA, and \ion{O}{iv} 4344 and
4389\,\AA) disappear at too-high temperatures, while the lines from
stage V disappear or become too weak at too-low temperatures (e.g.
the \ion{O}{v} multiplet at 4122\,\AA). In addition, the unobserved
\ion{O}{iii} multiplet around 3720\,\AA\ and the unobserved $3s-3p$
doublet of \ion{O}{vi} at 3811/3834\,\AA\ show up in models that are
too cool and too hot, respectively. Also, in too-cool models,
\ion{C}{iii} multiplets at 3885 and 4070\,\AA\ are seen, but they are
not present in the observed spectrum. Consequently, a good compromise
is achieved at \Teff = 80\,000\,K (and \logg = 6.0). Increasing and
decreasing \Teff\ by 5000\,K can roughly be compensated by
increasing and decreasing \logg\ by 0.5 dex and 0.3 dex, respectively,
such that the IV/V ionisation balances of nitrogen and oxygen hardly
change. What do change, however, are the profiles of the
\ion{He}{ii} lines and the \ion{C}{iv} lines. Accounting for the
uncertainties by blends of the unmodelled, highly excited \ion{C}{iv}
lines, we think that a good compromise is \logg = 6, but we assigned a
large conservative error of 0.5 dex.

The \ion{N}{v} 4945\,\AA\ multiplet in the final model is an
absorption line that is, however, not observed. In the case of \jl
presented below, we argue that this line is not a good diagnostic
tool because of the uncertainty in atomic data. The strength of the
\ion{N}{v} $3s-3p$ doublet (4604/4620\,\AA) of the same ion fits quite
well. Finally, we take a look at \ion{He}{i} 5876\,\AA. There
is a very weak feature in the observation, but it might not be
real. The line does not appear in our final model, but it would be too
strong at lower temperatures or higher gravities, which are outside of
our error range.

The radial velocity of \rl measured from the LAMOST spectrum amounts to
  \vrad = $+23\pm10$\,km/s and is listed in
  Table~\ref{tab:resultsall}.

\subsection{The O(He) star \jl}
\label{sect:helium}

\subsubsection{Effective temperature, surface gravity, and H/He ratio}
\label{sect:teff}

Again, the TMAP code was employed to construct model atmospheres. The
models include H, He, C, N, O, Si, P, S, Fe, and Ni. The employed
model atoms are described in detail by \citet{2018A&A...609A.107W}. In
addition, we performed line formation iterations (i.e. keeping the
atmospheric structure fixed) for argon using the model atom presented
in \citet{2007A&A...466..317W}.

The atmospheric parameters were inferred from the optical and UV
spectra. The employed lines are identified in Figs.\,\ref{fig:jl9_he},
\ref{fig:jl9_opt}, \ref{fig:jl9_hst}, and \ref{fig:jl9_fuse}. The
resulting values are listed in Table\,\ref{tab:resultsall}.

The effective temperature was constrained from elements exhibiting
lines from two or three ionisation stages. These are \ion{N}{iv} and
\ion{N}{v} lines in both the optical and UV, \ion{O}{iv} and
\ion{O}{v} lines in the optical plus the \ion{O}{vi} resonance doublet
in the FUSE spectrum, and \ion{Fe}{vi-viii} lines in the UV. The very
weak \ion{He}{i} 5876\,\AA\ line is also quite temperature
sensitive. The ionisation balances of these elements also depend on
the surface gravity, which at the same time is constrained by the
broad \ion{He}{ii} lines in the optical and the UV range. An iterative
procedure was performed to find the best fit model and to estimate the
error ranges. We arrived at  \Teff $= 80\,000 \pm 5000$\,K and \logg $
= 5.2 \pm 0.3$. The hydrogen abundance was determined from the
H$\alpha$ and H$\beta$ lines. It amounts to H =
$0.058^{+0.029}_{-0.019}$ (mass fraction).

\subsubsection{Metal abundances}
\label{sect:metals}

The carbon abundance was determined from the \ion{C}{iv} lines at 1108
and 1169\,\AA\ in the FUSE spectrum. For nitrogen, several lines of
\ion{N}{iv} and \ion{N}{v} in the optical as well as in the FUSE and
HST spectra were used. \ion{O}{iv} and \ion{O}{v} lines in the optical
and the \ion{O}{vi} resonance doublet in the FUSE range served to
derive the oxygen abundance.

Other light metal abundances were exclusively constrained from lines
in the FUSE spectrum. Fluorine can be studied by the \ion{F}{vi} line
at 1139.5\,\AA. However, its identification in \jl is uncertain. For
silicon, phosphorus, and sulphur we used the lines of \ion{Si}{iv} at
1122.5 and 1128.3\AA, \ion{P}{v} at 1118.0 and 1128.0\,\AA, and
\ion{S}{vi} at 933.4, 944.5, 1000.5 and 1117.8\,\AA. The argon
abundance was found from the \ion{Ar}{vii} line at 1063.5\,\AA. The very large number of lines of \ion{Fe}{vi-viii} and \ion{Ni}{vi}
in the FUSE and HST spectra were used to determine the iron and nickel
abundances.

\begin{figure}[t]
 \centering  \includegraphics[width=0.94\columnwidth]{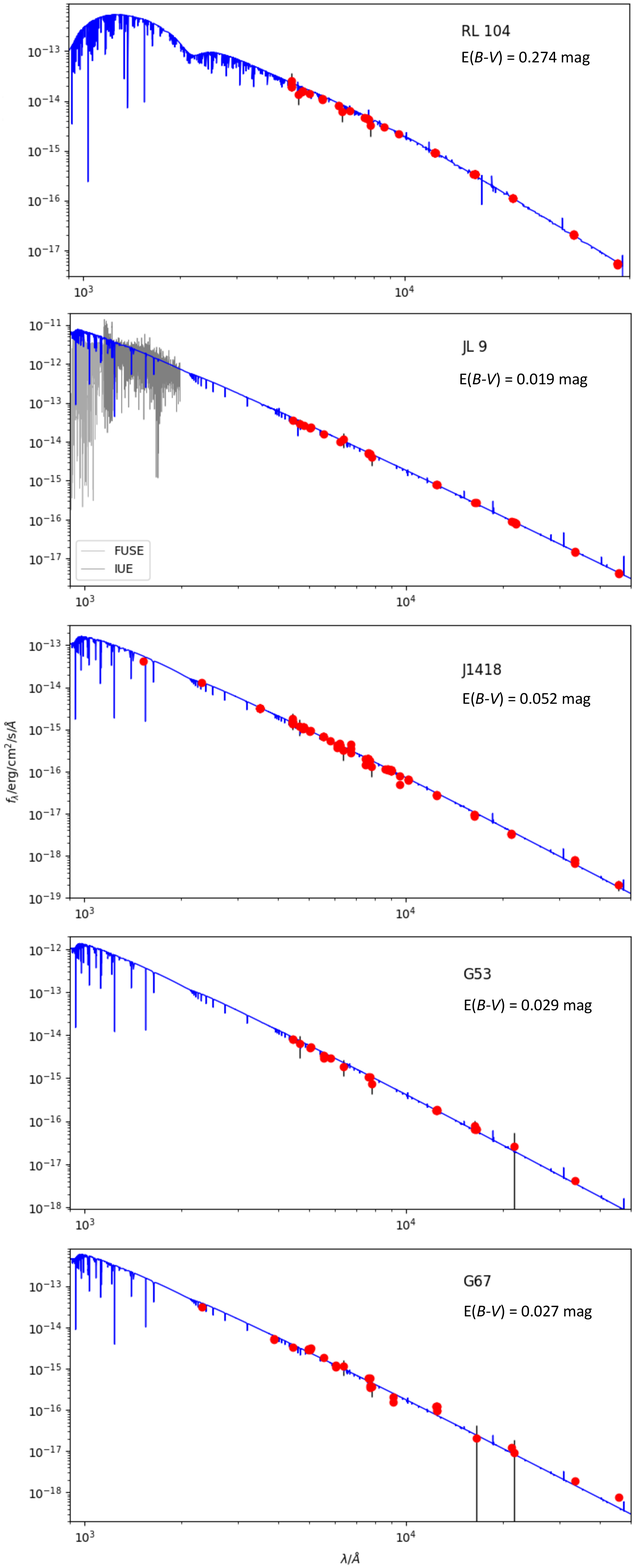}
     \caption{Comparison of our best fitting model fluxes (blue) with
       the observed photometry (red; in the case of \jl, the FUSE and
       IUE spectra are shown in grey). The reddening that has been
       applied to the model fluxes is indicated in each panel.}
    \label{fig:sed}
\end{figure}

\begin{figure*}[t]
 \centering
 \includegraphics[width=0.9\textwidth]{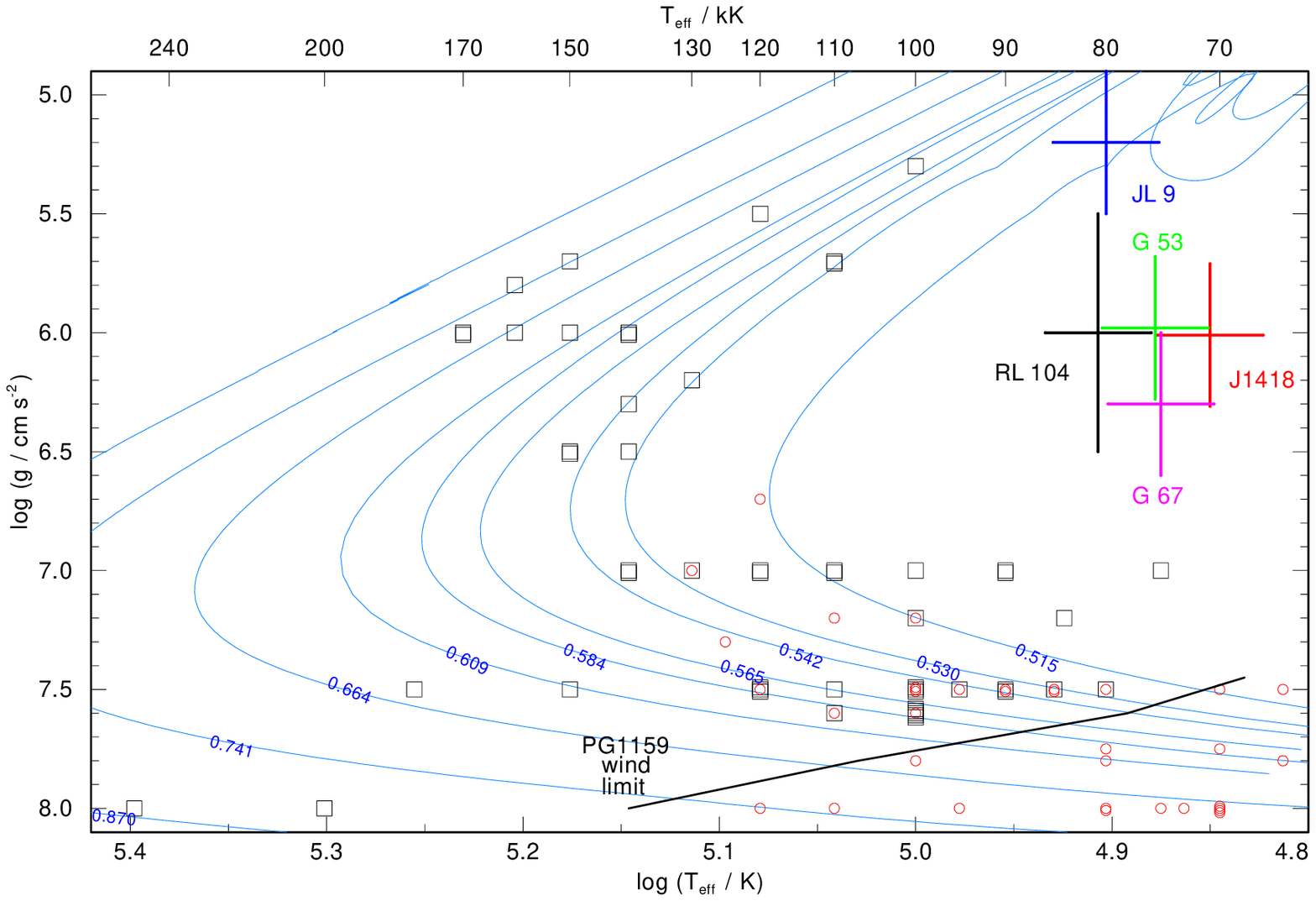}
  \caption{Position of our programme stars in the
    \Teff--\logg\ diagram, in comparison with other PG1159 stars
    (black squares) and DO white dwarfs (red circles). Blue lines are
    VLTP post-AGB evolutionary tracks (labelled with stellar mass in
    $M_\odot$) from \cite{2006A&A...454..845M}. Indicated is the
    PG1159 wind limit according to \cite{2000A&A...359.1042U}. No PG\,1159 stars are found below it because radiation-driven stellar
    winds become so weak that they can no longer prevent gravitational
    settling of heavy elements.}
\label{fig:pg1159_evolution}
\end{figure*}

\subsubsection{Problems with the \ion{N}{v} emission line at 4945\,\AA}

In Fig.\,\ref{fig:jl9_opt}, we note a significant problem in fitting the
\ion{N}{v} emission line at 4945\,\AA. The line is a blend of three
transitions between levels with principal quantum numbers $n=6$ and 7,
namely, $6f-7g$, $6g-7h$, and $6h-7i$. In Fig.\,\ref{fig:jl9_nv4945},
we display the situation in detail. The wavelength position of the
line blend stems from a laboratory experiment by \cite{Hallin1966}, in
which this blend was not resolved either. He calculated the positions of
these three components (indicated in Fig.\,\ref{fig:jl9_nv4945} and in
Table\,\ref{tab:nv}). In contrast, if we calculate the components'
positions that follow from level energies listed in the NIST database,
they are rather different. The separation is wider, and we employed
these positions usually for our spectrum synthesis
calculations. Obviously it is too wide because they are inconsistent
with the spectrum of \jl. We observe an asymmetry in the observed line
profile such that an emission bump appears in the blue wing, which
probably stems from the $6f-7g$ component whose position seems to be
in agreement with the calculation by \cite{Hallin1966}. \cite{1995PhyS...52..166R} reassessed this line blend in laboratory
spectra and managed to resolve the $6f-7g$ component from the others (and
assigned a wavelength to it that was consistent with Hallin's
calculation), but still the two other components were not resolved. In
order to study what could be the effect of a narrower separation of
the line components, we chose the other extreme and shifted all three
of them to one and the same wavelength position \citep[4944.46\,\AA,
  as measured by][]{Hallin1966} and carried out the following numerical
test. The line overlap was considered in the NLTE iterations for the
population numbers and in the final formal solution for the spectrum
synthesis in a simplified model atmosphere (less elements, \Teff =
80\,000, \logg = 5.2). The result is shown in
Fig.\,\ref{fig:jl9_opt}. The absorption wings  visible in the original
model (with NIST splittings) disappear and the central emission at
4944.46\,\AA\ is quite pronounced, although the line is not wide
enough compared to the observation. A likely explanation is the
presence of a weak wind. We conclude that the use of this line as a
diagnostic tool is not possible until more precise wavelength
measurements are performed in laboratory.

The radial velocity of \jl measured from the UVES spectrum is \vrad =
$+76.4\pm1.0$\,km/s. It agrees with the value of $+77\pm2$\,km/s
measured by \citet{1987fbs..conf..603D} from the very same CASPEC
spectrum that we also used for our spectral analysis.

\subsection{The He-sdOs \sdsss, \gaiaff, and \gaiass}
\label{sect:helium2}

The analysis of the optical spectra proceeded in analogy to \jl. For
\sdsss we found \Teff $= 70\,000 \pm 5000$\,K and \logg $ = 6.0 \pm
0.3$. The presence of \ion{He}{i} lines helped to constrain the
temperature together with the observed lines of C, N, and O (or their
absence), while the gravity is indicated by the \ion{He}{ii} lines.
For H, He, C, N, and O we measured the abundance values given in
Table~\ref{tab:resultsall}. The errors for C on one hand and N and O
on the other hand were estimated to $\pm$\,0.2 and $\pm$\,0.3\,dex,
respectively. In the model calculations, the mass fractions for Si, P,
S, Fe, and Ni were set to solar. UV spectra would be needed to assess
their abundances. The model fit is presented in
Fig.\,\ref{fig:sdssj141812_opt}.

\gaiaff turned out to be a twin of \sdsss, except for the effective
temperature (see Table\,\ref{tab:resultsall}). For \gaiaff it is
higher by 5000\,K because the \ion{He}{i} lines at 4471\,\AA\ and
5876\,\AA\ are weaker. The O abundance can be determined from the very
weak \ion{O}{iv} triplet at 3730\,\AA.  The model fit is presented in
Fig.\,\ref{fig:gaiaf_opt}.

\gaiass has the same temperature but a slightly higher gravity
(75\,000\,K, \logg = 6.3). The main difference to the other He-sdO
stars is that carbon cannot be detected. The \ion{O}{iv} triplet at
3730\,\AA\ is not identified, perhaps because of the lower signal-to-noise ratio of the
spectrum. Therefore, only upper limits for C and O were derived.

We measured the radial velocities of the three He-sdOs and list
  them in Table~\ref{tab:resultsall}. The value for \sdsss (\vrad =
  $-55\pm3$\,km/s) is in agreement with the one we measured
  from the SDSS spectrum ($-55\pm10$\,km/s) that was analysed by
  \cite{2014A&A...564A..53W}.\footnote{We note that the SDSS spectra
    analysed by \cite{2014A&A...564A..53W} were shifted from vacuum to
    rest wavelengths by amounts given in their Table~3.} 

\begin{figure*}[t]
 \centering  \includegraphics[width=0.9\textwidth]{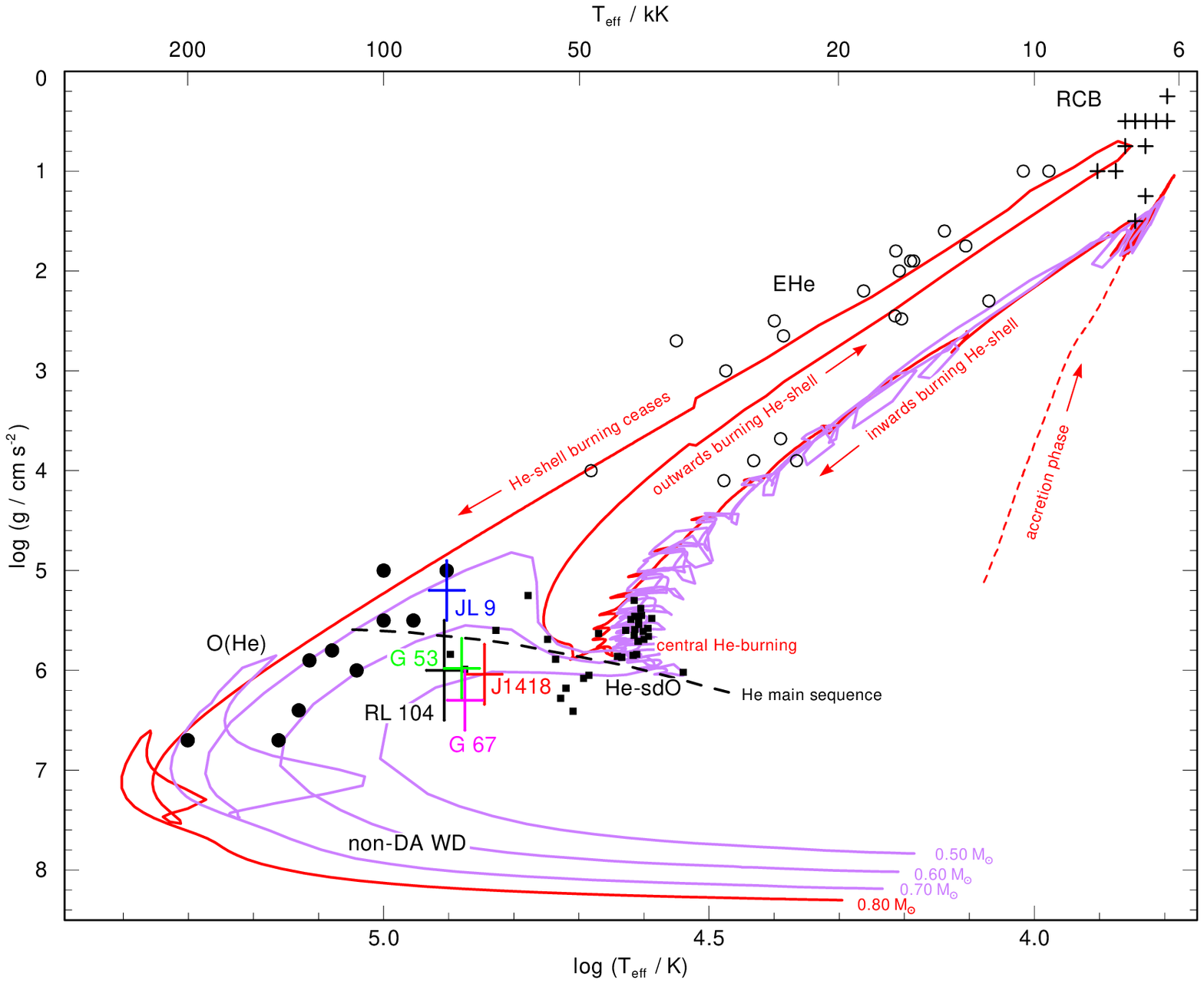}
  \caption{Position of our programme stars in the
    \Teff--\logg\ diagram among other O(He) stars (full circles) as
    well as RCB stars (crosses), EHe stars (open circles), and He-sdOs
    (squares), see \citet{2014A&A...566A.116R} and references
    therein. Red and purple lines are evolutionary tracks for a
    composite-merger scenario of two He-WDs (marked with the merger
    mass in $M_\odot$) from \cite{2012MNRAS.419..452Z}. The dashed
    black line is the zero-age helium main sequence.}
\label{fig:merger_evolution}
\end{figure*}

\section{Physical parameters}
\label{sect:sed}

\subsection{SED fitting, stellar radii, luminosities, and distances}

With well-known distances, $d$, from the Gaia eDR3 \citep{Gaia+2020}, it
is now possible to determine the radius, $R$, luminosity, $L$, and
mass, $M$, of a star, if its effective temperature and surface gravity
are known from spectroscopy. For that, one needs to perform a fit to
the observed SED which takes into account the effect of interstellar
reddening. The observed flux, $f_\lambda$, relates to the (reddened)
model flux, $F_\lambda$, via $f_\lambda=F_\lambda\pi R^2/d^2$. In
order to derive the radii of our stars, we used the distances reported
by \citet{2021AJ....161..147B} and then (using the
\cite{Fitzpatrick1999} reddening law) varied the values for \ebv and
the radius, until a good agreement of our best fitting model fluxes
from our spectral analysis to the observed SED was found. We employed
photometry from various catalogues: GALEX \citep{Bianchi+2017},
Pan-STARRS1 \citep{Flewelling+2020}, Landolt B, V \citep{Henden+2015},
Gaia \citep{Gaia+2020}, SkyMapper \citep{2018PASA...35...10W}, SDSS
\citep{2015ApJS..219...12A}, 2MASS \citep{Cutri2003}, and WISE
\citep{Schlafly+2019}. Magnitudes were converted into fluxes using the
VizieR Photometry
viewer\footnote{\url{http://vizier.unistra.fr/vizier/sed/}}. In the case
of \jl, the FUSE and IUE spectra were also employed for the fit. The
resulting values for the reddening and radii are listed in
Table\,\ref{tab:resultsall}. This table also lists the luminosities of
our stars that were calculated from the radii and effective
temperatures via $L/L_\odot =
(R/R_\odot)^2(T_\mathrm{eff}/T_{\mathrm{eff},\odot})^4$.  In
Fig.\,\ref{fig:sed}, our SED fits are shown. It can be seen that the
SEDs of all our five stars are reproduced nicely and there is no sign
of a companion or a disk.

To evaluate the spectroscopic distances of the stars, we must consider
the effect of interstellar reddening. The visual extinction is
evaluated from the standard relation $A_V=3.1 \times$ \ebv, so we
obtain a dereddened visual magnitude $V_0$. The spectroscopic distance
$d$ is found by the relation
$$ d {\rm [pc]}= 7.11\times 10^{4} \sqrt{H_\nu\cdot M\cdot 10^{0.4
    V_0-\log g}}\ ,$$
where $H_\nu$ is the Eddington flux of the best-fit atmosphere model
at 5400\,\AA\ (Table~\ref{tab:resultsall}). The results are listed in
Table~\ref{tab:resultsall} as well as the distances derived from the
Gaia EDR3 parallaxes \citep{2021AJ....161..147B}. The main error
source for the spectroscopic distance is the uncertainty in the
surface gravity. Both distance determinations agree within error
limits except for \gaiass for which the parallax distance is a factor
of two larger. We note that similar problems were encountered and
discussed, for instance, in spectroscopic analyses of hot WDs
\citep{2019MNRAS.487.3470P} and hot post-AGB stars
\citep{2020MNRAS.494.2117H}.

\subsection{Stellar masses}

Figures\,\ref{fig:pg1159_evolution} and \ref{fig:merger_evolution} show
the position of the analysed stars in the \Teff--\logg\ diagram. Their
masses are determined by comparison with evolutionary tracks by linear
interpolation or extrapolation. The errors are dominated by the
uncertainty in $g$. From Fig.\,\ref{fig:pg1159_evolution}, we find the
mass of the PG1159 star \rl\ by comparison with tracks for stars that
experienced a very late thermal pulse (VLTP)
\citep{2006A&A...454..845M}. It is $M=0.48^{+0.03}_{-0.02} M_\odot$.
For the O(He) star \jl, these tracks imply $M=0.52^{+0.09}_{-0.02}
M_\odot$. Alternatively, the stellar merger tracks in
Fig.\,\ref{fig:merger_evolution} yield $M=0.68^{+0.11}_{-0.06}
M_\odot$. For the He-sdOs, the masses are derived from the merger
tracks and listed in Table\,\ref{tab:resultsall}.

Because the distances are well known by the Gaia parallax
measurements, the stellar masses can be determined in an alternative
way. From the SED fit that used the Gaia distance, we know the stellar
radius $R$. Together with the spectroscopically determined surface
gravity $g,$ we obtain the mass from $M/M_\odot = g/g_\odot \times
(R/R_\odot)^2$. However, the large uncertainty in $g$ (factors 3.2 and
2 for \rl and the other for stars, respectively) directly propagate
into the mass determination. For \rl, \jl, \sdsss, \gaiaff, and
\gaiass, we find $M/M_\odot$ = $1.31^{+3.75}_{-1.02}$,
$0.36^{+0.61}_{-0.23}$, $0.36^{+0.86}_{-0.27}$,
$0.36^{+0.68}_{-0.24}$, and $1.62^{+4.16}_{-1.26}$, respectively. The
values are formally consistent with the masses found from the
evolutionary tracks, but cannot further constrain them.

\section{Discussion and summary}
\label{sect:discussion}

\begin{figure*}[t]
 \centering  \includegraphics[width=0.9\textwidth]{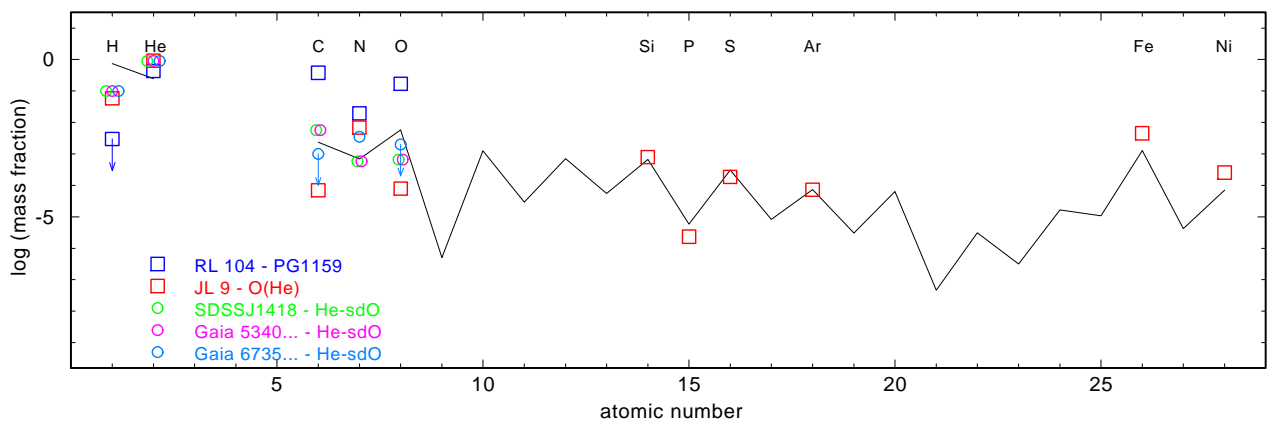}
  \caption{Abundances measured in our programme stars. Arrows indicate
    upper limits. The black line represents solar values.}
\label{fig:abu}
\end{figure*}

Our three analysed He-sdOs are located below the helium main sequence
and we identify them with evolved, He-shell burning post-EHB stars
that either formed from a binary WD merger or by single star evolution
with a late helium core flash. The O(He) star \jl is a more luminous
object that lies in the post-AGB region of the Kiel diagram. The
abundance pattern of O(He) stars cannot be explained by (V)LTP models
\citep{2014A&A...566A.116R}, and hence we discuss its evolution in the
context of a binary He-WD merger. The PG1159 star \rl is discussed as
a (V)LTP object, as is usual for this stars in this spectral class
\citep{2006PASP..118..183W}. However, in light of its low mass, a binary WD
merger model is also considered. The different evolutionary scenarios
are discussed with regard to determined abundance patterns that are
displayed in Fig.\,\ref{fig:abu}.

\subsection{Evolution of the O(He) star \jl and the three He-sdOs}

The He/H number abundance ratio in our three He-sdOs is $\log y =
0.35$. This is, for example, in good agreement with the values found for many
hot He-sdOs, including those in the GALEX sample
\citep{2012MNRAS.427.2180N}, which is around $\log y = 0.6$
\citep{2016PASP..128h2001H}.

As for the C and N abundances, the abundance patterns of our He-sdOs
are similar to other stars of this class. In \gaiass, we see a clear
hint to CNO burning. N is strongly enriched compared to the Sun, while
C and O are depleted. In \sdsss and \gaiaff, C is enriched by
He-burning while O is depleted. \citet{2010AIPC.1314...79H} showed
that in their ESO/SPY sample all objects with \Teff $>$ 43\,000\,K are
carbon dominated (C/N $>$ 1, by mass; see also
\citet{2007A&A...462..269S} and \citet{2018A&A...620A..36S} ). In stark
contrast to these 15 objects, \gaiass is nitrogen dominated (N/C $>$
3.5). Oxygen abundance measurements in He-sdOs are rather scarce. For
four objects with \Teff\ around 45\,000\,K,
\citet{2018A&A...620A..36S} found logarithmic mass fractions between
$-4.2$ and $-2.9$. For \sdsss and \gaiaff we determined an
intermediate value of $-3.2$ and an upper limit of $-2.7$ for
\gaiass. For the O(He) star we found a strong N enrichment ($-2.2$)
whereas C and O are strongly depleted ($-4.2$ and $-4.1$,
respectively).

We compare the abundances of our He-sdOs and the O(He) star to surface
abundance predictions by the binary He-WD merger models of
\citet{2012MNRAS.426L..81Z}. They comprise different merger
scenarios, a slow (cool) merger, a fast (hot) merger, and a
combination of both (composite merger). In the first case, the merger
product exhibits the composition of the accreted WD, that is He-rich
matter as CNO burning ash with C and O transformed into N. In the
second case, nuclear burning and mixing changes the abundances in
favor of He-burning ashes, creating C overabundances and depleting
N. The composite merger combines both processes. In this case, mergers
with $M<0.7M_\odot$ are N rich, while more massive ones are C rich
with significant N abundance.

For the He-sdO \gaiass and the O(He) star \jl, we have N oversolar and
C and O depleted; hence, this fits the cold-merger model. CNO
abundances of three of the four O(He) stars analysed by
\citet{2014A&A...572A.117R} show a similar behaviour. The high C/N =
10 mass ratio in \sdsss and \gaiaff, however, requires
composite-merger models (because of the presence of N, a fast-merger
model does not fit) with high merger mass (0.8\,$M_\odot$), because
they have efficient dredge up of C-rich material from the stellar
core. However, this is at odds with the spectroscopic masses
$M=0.50^{+0.06}_{-0.06} M_\odot$ and $M=0.53^{+0.07}_{-0.05}
M_\odot$. Another disaccord is the relatively high hydrogen abundance
that we found in the O(He) star and the He-sdOs (6 and 10\% by
mass). The merger models predict much lower abundances
\citep{2016MNRAS.463.2756H,2018MNRAS.476.5303S,2020JApA...41...48J}. The
abundances of heavier elements in \jl (Si, P, S, Ar, Fe, and Ni) are
essentially solar, reflecting the original composition of the star.

The late hot flasher scenario is the alternative explanation for
the origin of He-sdOs. Usually, a helium-core flash occurs at the tip
of the red giant branch (RGB). However, this flash can also occur when
a star leaves the RGB to become a WD or even when it is a WD. In the
latter case, the star is termed a late hot flasher, and the
flash-driven convection reaches the surface layer altering the
chemical composition. The scenario has two types, namely a shallow
mixing (SM) case and deep mixing (DM) case. In the first case,
hydrogen is ingested into the stellar interior and diluted such that
the surface abundance of H is reduced, while in the second case H is
ingested an burned and only minute quantities of H remain visible on
the surface. We compared the surface abundances predicted by models
for late hot flashers from \citet{2018A&A...614A.136B} with those of
our three analysed He-sdOs. For that, we chose the model with
metallicity $Z=0.02$ and initial helium abundance He = 0.285. One
immediately faces the problem that, in the SM case, the predicted
hydrogen abundance (H = 0.47) is much higher than the observations (H
= 0.10), and in the DM case the prediction is much lower (0.0036
or even smaller, depending on details in the modelling). The SM model
also predicts C four times lower and N ten times higher than observed
for \sdsss and \gaiaff. For \gaiaff, the predicted nitrogen abundance
(N = 0.0062) agrees with the observed one (N = 0.0035) within error
limits (0.5~dex), just like the predicted C and N abundances (C =
0.0015, O = 0.0062) with the upper limits from the observations (C $<$
0.001, O $<$ 0.002). For the O(He) star \jl, the observed C and O
abundances are by more than 1~dex below the hot-flasher prediction
and, like for the He-sdOs, the observed hydrogen abundance (H = 0.058)
cannot be fitted by the DM and SM models.

In essence, the main obstacle for both, the merger model and the hot
flasher model is the failure to explain the observed hydrogen
abundance in our He-sdOs. The O(He) star's CNO abundances are in
accordance with the merger model, but not its high hydrogen content.

In this context, we note that \citet{2017ApJ...835..242Z} pointed out
that hot subdwarfs with intermediate H and He abundances (He number
fractions 10\%--90\%) can be formed by mergers of a main-sequence star
with a helium WD. However, their models predict that the atmospheres
of the merger remnants become pure hydrogen even before they reach
the helium main sequence because of the gravitational settling of helium
and metals.

\subsection{Evolution of the PG1159 star \rl}

The atmospheric composition of \rl (He = 0.43, C = 0.38, N = 0.02, O =
0.17, mass fractions) is typical for PG1159 stars. The prototype
PG1159$-$035, for instance, has He = 0.35, C = 0.48, N = 0.001, and O = 0.17
\citep{2006PASP..118..183W}. Using the VLTP tracks, we derived a mass
of $M=0.48^{+0.03}_{-0.02} M_\odot$ for \rl. It is one of the lowest
masses found for any PG1159 star, beaten only by HS\,$0704+6153$,
which has a comparably low mass of $M=0.47 M_\odot$
\citep{1998A&A...334..618D,2006A&A...454..845M}. The difference
between the two objects is, however, that \rl\ is a He-shell burner
evolving towards higher effective temperatures, while HS\,$0704+6153$
is already on the WD cooling track (\logg = 7.5), it being the coolest
PG1159 star to date (\Teff = 75\,000\,K). Therefore, \rl\ is the
coolest PG1159 star (\Teff = 80\,000\,K) hitherto found in a pre-WD
evolutionary stage (\logg = 6.0). The presence of nitrogen is an
indicator that the star suffered a very late thermal pulse and not
another variant, namely a late thermal pulse or an AGB final thermal
pulse (AFTP). For a detailed discussion of these scenarios we invite the reader to consult, for example,
\citet{2006PASP..118..183W}.

From the fact that \rl is located among the hottest He-sdOs in the
\logg--\Teff diagram, it is tempting to consider that it is the
outcome of a binary WD merger. However, the binary merger models of
two He-WDs of \citet{2012MNRAS.426L..81Z} predict helium-dominated
envelopes and cannot explain the high-C and -O content in the
atmosphere of \rl. The situation is different when both merging WDs
have a C/O core. Recently, \citet{2021ApJ...920..110S} presented
calculations in order to predict the outcome of a close binary system
consisting of a hot subdwarf and a WD, which is a new class of
Roche-lobe filling binaries where the subdwarfs transfer mass to the
WD companion \citep{2020ApJ...898L..25K,2020ApJ...891...45K}. The
subdwarf will evolve into a low-mass C/O core WD with a thick helium
layer, and it will finally merge with the more massive C/O
WD. \citet{2021ApJ...920..110S} followed the post-merger evolution
after constructing an initial model in which a $0.55\,M_\odot$ C/O WD
accreted matter from the $0.36\,M_\odot$ C/O remnant of a He-sdO. In
the pre-WD phase (steady He-shell burning) of the merger, the surface
abundances are about He = 0.30, C = 0.26, N = 0.004, and O =
0.44. Considering the error limits of our abundance determination in
\rl (0.3\,dex), the abundance pattern from this model is similar to
the observation. The very low hydrogen abundance H = 0.001 in the
merger model is in accordance with our detection threshold of H $<$
0.003. The high mass of the merger model ($\approx 0.9\,M_\odot$) is,
however, at odds with that of \rl\ derived from the merger models of
\citet{2012MNRAS.426L..81Z} in Fig.\,\ref{fig:merger_evolution}
($M=0.53^{+0.10}_{-0.09} M_\odot$).

Just before reaching the WD sequence, the high-mass He-WD$+$He-WD
merger model sequences (0.7 and $0.8\,M_\odot$) of
\citet{2012MNRAS.426L..81Z} display a small loop caused by a final
He-shell flash (Fig.\,\ref{fig:merger_evolution}). One can speculate
that it initiates envelope convection, creating PG1159-like abundances
similar to the VLTP scenario in single star evolution. Maybe with
enhanced physical models, such a late He-shell flash could also occur
in lower mass mergers (Josiah Schwab, priv. comm.), explaining the
existence of low-mass PG1159 stars such as \rl.

\begin{acknowledgements} 
We thank Uli Heber for putting his CASPEC spectrum of \jl\ at our
disposal and useful comments on an earlier version of the manuscript,
and Max Pritzkuleit for his support of the HOTFUSS project.
M.D.\ acknowledges funding by the Deutsche For\-schungs\-gemeinschaft
(DFG) through grant HE1356/71-1. R.R.\ has received funding from the
postdoctoral fellowship programme Beatriu de Pin\'os, funded by the
Secretary of Universities and Research (Government of Catalonia) and
by the Horizon 2020 programme of research and innovation of the
European Union under the Maria Sk\l{}odowska-Curie grant agreement No
801370.  The TMAD tool (\url{http://astro.uni-tuebingen.de/~TMAD})
used for this paper was constructed as part of the activities of the
German Astrophysical Virtual Observatory. Some of the data presented
in this paper were obtained from the Mikulski Archive for Space
Telescopes (MAST). This research has made use of NASA's Astrophysics
Data System and the SIMBAD database, operated at CDS, Strasbourg,
France. This research has made use of the VizieR catalogue access
tool, CDS, Strasbourg, France. This work has made use of data from the
European Space Agency (ESA) mission Gaia.

\end{acknowledgements}

\bibliographystyle{aa}
\bibliography{aa}

\end{document}